\documentclass{appolb}

\usepackage{amssymb}
\usepackage{amsmath}
\usepackage{epsfig}

\newcommand{\be}{\begin{equation}}
\newcommand{\ee}{\end{equation}}
\newcommand{\bea}{\begin{eqnarray}}
\newcommand{\eea}{\end{eqnarray}}

\begin{document}


\title{Diffractive and photon-induced processes at the LHC: from the odderon discovery, the evidence for saturation  to the search for axion-like particles}

\author{Christophe Royon 
\address{The University of Kansas, Lawrence, USA \\
email: christophe.royon@cern.ch} }


\maketitle

\begin{abstract}
We discuss first the discovery of the odderon by the TOTEM and D0 collaborations. We then describe the gap between jets measurements sensitive to the high gluon density regime and the possible observation of saturation phenomenon in Pb Pb interactions. We also mention the sensitivity to beyond standard model physics and to the production of axion-like particles via photon photon interactions.
\end{abstract}

In this paper, we present some recent results on Quantum Chromodynamics (QCD) diffraction at the LHC, mainly from the CMS and TOTEM experiments. We start by presenting the discovery of the odderon by comparing the elastic cross section measurements in $pp$ and $p \bar{p}$ interactions by the TOTEM and D0 collaborations at the LHC and the Tevatron. We then discuss the possible observation of the high gluonic density regime at the LHC, and especially of saturation phenomenon in heavy ion interactions such as Pb Pb. We finish this report by discussing the sensitivity to beyond standard model physics and the production of axion-like particles considering the LHC as a $\gamma \gamma$ collider.

These studies originate from 
a long term collaboration with Prof. Andrzej Bialas and Prof. Robi Peschanski 
that started after my PhD in Saclay about the dipole model and
diffraction~\cite{dipole} and I would like to express all my gratitude to Andrzej and Robi
for this long term and successful collaboration, and I wish him a very nice
birthday.

\section{The discovery of the odderon}

\begin{figure}
\begin{center}
\epsfig{file=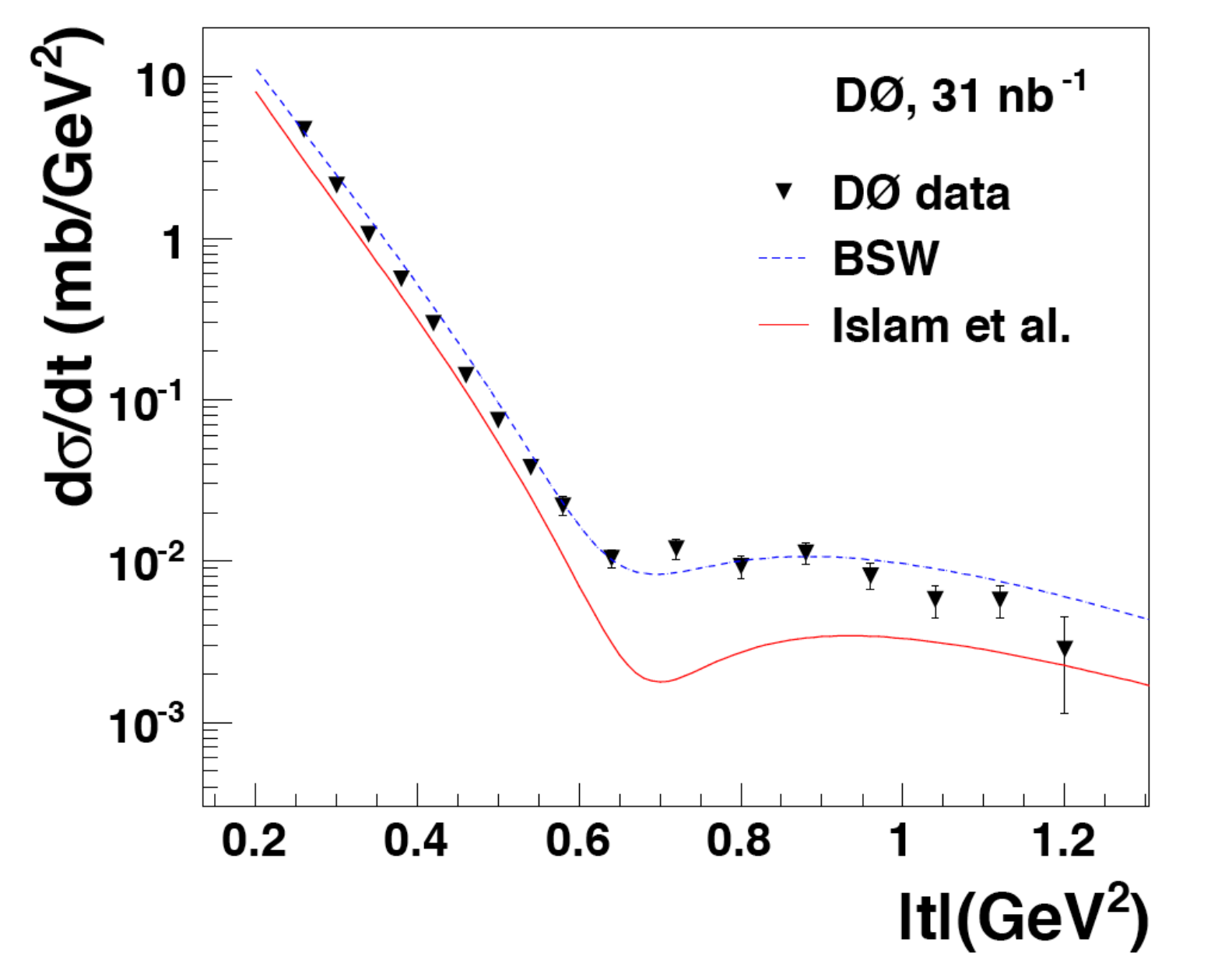,width=8.5cm}
\caption{$p \bar{p}$ elastic $d \sigma/dt$ cross section as a function of $|t|$ measured by the D0 collaboration at 1.96 TeV}
\label{d0}
\end{center}
\end{figure}

\begin{figure}
\begin{center}
\epsfig{file=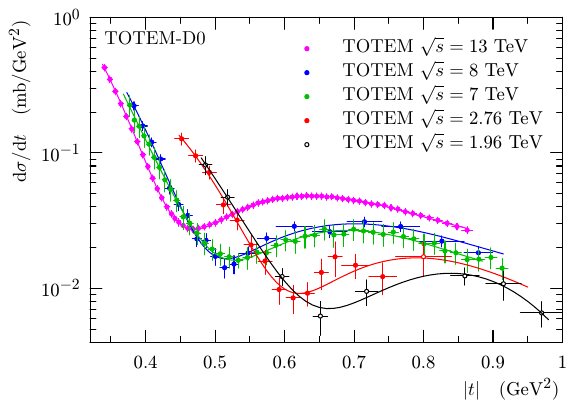,width=8.5cm}
\caption{$pp$ elastic $d \sigma/dt$ cross section as a function of $|t|$ measured by the TOTEM collaboration at 2.76, 7, 8 and 13 TeV and extrapolated down to 1.96 TeV in black.}
\label{totem}
\end{center}
\end{figure}

The comparison of the elastic event measurements at the LHC and the Tevatron, respectively in $pp$ and $p \bar{p}$ collisions led recently to the discovery of the odderon~\cite{ourpaper}. In elastic interactions,  each proton or antiproton remains intact after interaction and is scattered at some angles and can lose or gain some momentum. Nothing is produced in addition to the intact proton or antiproton. Measuring elastic collisions thus requires the detection of intact protons or anti-protons after the collision, using dedicated detectors called roman pots. 
These detectors can move very close to the beam (up to 3$\sigma$ at the LHC for instance) when beam are stable so that protons scattered at very small angles can be measured. The TOTEM~\cite{totem} and ATLAS/ALFA~\cite{alfa} collaborations installed roman pots at about 220 m from the interaction point and the D0 collaboration had similar detectors at about 23 and 31 m~\cite{FPD}. Just before the end of the Tevatron, the D0 collaboration measured the elastic $p\bar{p}$ $d\sigma/dt$ cross section at  a center-of mass energy $\sqrt{s}$ of 1.96 TeV for $0.26<|t|<1.2$ GeV$^2$ where $t$ is the quadri-momentum transferred square
at the proton/antiproton vertex measured by tracking the proton/antiproton~\cite{d0cross} as displayed in Fig.~\ref{d0}. The TOTEM collaboration measured the elastic $pp$ $d\sigma/dt$ at $\sqrt{s}=$2.76, 7, 8 and 13 TeV~\cite{totemdata} as shown in Fig.~\ref{totem}. The advantage of the TOTEM detector is that it includes very forward telescope covering a rapidity region $3.1<|\eta|<4.7$ (telescope T1) and $5.3<|\eta|<6.5$ (telescope T2), allowing to veto an additional particles emitted in the very forward region for elastic interactions. 

Elastic scattering can be explained by the exchange of colorless objects, pomeron and odderon, that
correspond to positive and negative charge parity $C$~\cite{nicolescu,martynov,landshoff,leader}. The odderon is defined as a singularity in the complex plane of the $t$-channel partial wave $J$, located at $J = 1$, when $t = 0$ and which contributes to the odd crossing amplitude. The QCD treatment of the odderon was performed considering multi-gluon exchanges in hadron-hadron interactions in elastic $pp$ interactions~\cite{bkp}.  In QCD, the pomeron is made of an even number of gluons leading to a ($+1$) $C$-parity whereas the odderon is made of an odd number  of gluons corresponding to a ($-1$) $C$-parity.  
The elastic scattering amplitudes for $pp$ and $p \bar{p}$ interactions can be written as the sum or the difference of the even and the odd part of the amplitude
\begin{eqnarray}
A_{pp} &=& Even~+~Odd \\
A_{p \bar{p}} &=& Even~-~Odd .
\end{eqnarray}
From these equations, it is clear that observing a difference between $pp$ and $p \bar{p}$ interactions would be a clear way to observe the odderon. Our idea was thus to obtain a quantitative measurement of the differences between elastic $pp$ and $\bar{p} p$ cross sections.

\begin{figure}
\begin{center}
\epsfig{file=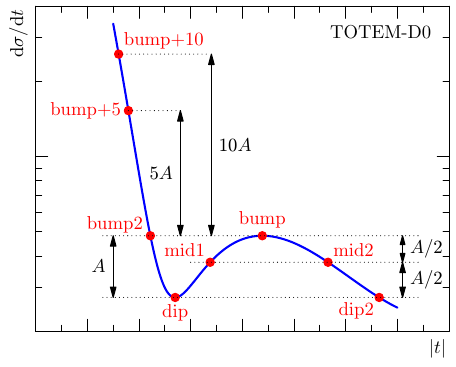,width=8.5cm}
\caption{Definition of the points that are characteristic of the shape of elastic $pp$ cross sections (see text).}
\label{reference}
\end{center}
\end{figure}

\begin{figure}
\begin{center}
\epsfig{file=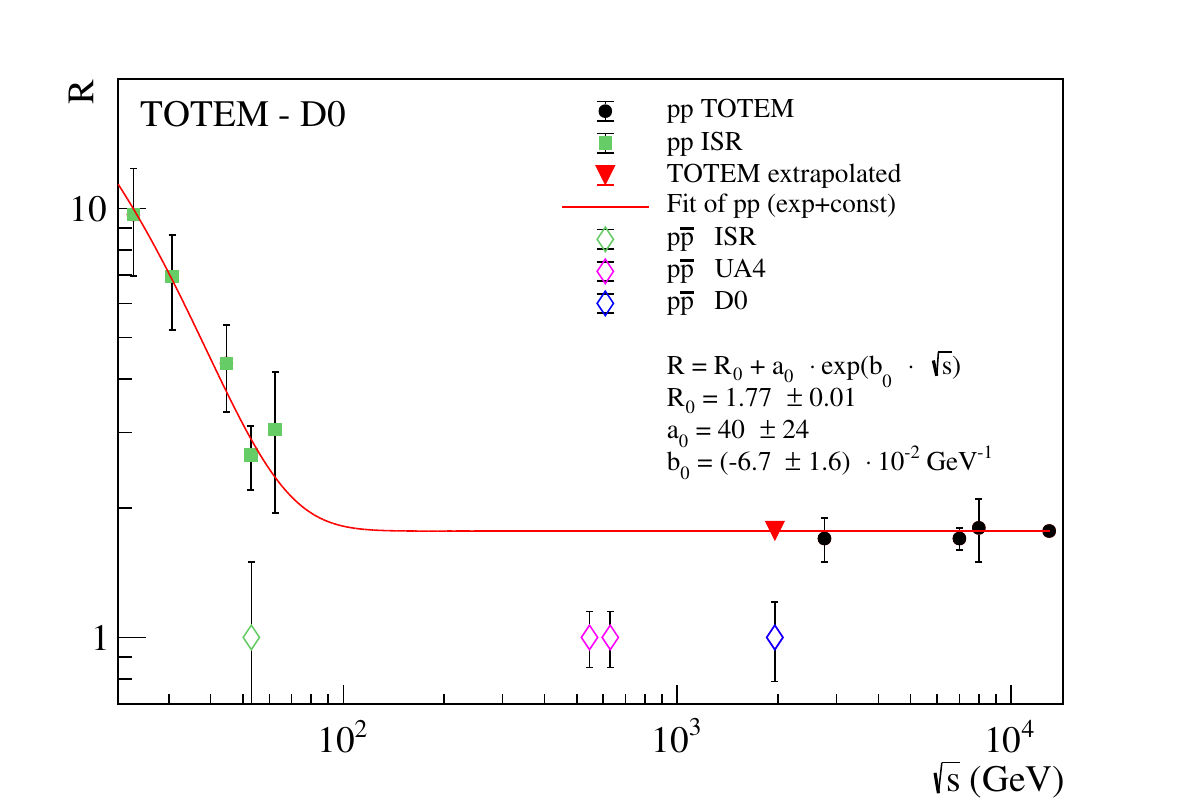,width=10.5cm}
\caption{Bump over dip ratio for $pp$ and $p \bar{p}$ elastic cross sections.}
\label{bumpoverdip}
\end{center}
\end{figure}

Let us note that there was already some indication of possible differences at ISR energies between $pp$ and $p \bar{p}$ interactions~\cite{isr} 
but this was not considered to be a proof of the existence of the odderon because of possible
additional reggeon and meson exchanges at lower $\sqrt{s}$. At high energies of the order of 1 TeV or higher, the contribution of reggeon and meson exchanges becomes negligible and a potential difference between $pp$ and $p \bar{p}$ elastic cross section is due to the odderon.

As seen in Fig.~\ref{d0} and \ref{totem}, elastic $d\sigma/dt$ cross sections show different behaviors between $pp$ and $p \bar{p}$ interactions. There is a maximum (the bump) and a minimum (the dip) for $pp$ interactions at all $\sqrt{s}$ whereas such behavior is not observed for $p \bar{p}$ interactions.
In order to compare directly elastic $pp$ and $p \bar{p}$ cross sections, one needs to extrapolate down the TOTEM measurements at 2.76, 7, 8 and 12 TeV  down to 1.96 TeV, the Tevatron $\sqrt{s}$. Unfortunately, a direct comparison is not possible since there is no acceptance for the TOTEM roman pots at 1.96 TeV, in the dip and  bump region where we perform the comparison. We define a series of eight points defined in the $d\sigma/dt$ cross section (we give both the $t$ and $d\sigma/dt$ values)  that are characteristic of the elastic $pp$ cross section shape as shown in Fig.~\ref{reference}, namely the $bump$, the $dip$. $dip2$ and $bump2$ corresponding to the same cross section as the dip and the bump but at lower $|t|$, $mid1$ and $mid2$ corresponding to the middle in cross section between the dip and the bump, and finally $bump+5$ and $bump+10$ that correspond to the cross sections 5 and 10 times the difference between the bump and the dip. 

The bump over dip ratio measured for $pp$ and $p \bar{p}$ elastic interactions at ISR~\cite{isr}, Tevatron and LHC energies
is shown in Fig.~\ref{bumpoverdip}. In $pp$ elastic collisions, it decreases as a function of $\sqrt{s}$ up to $\sim$ 100 GeV and is flat above. D0 $p \bar{p}$  shows a ratio of 1.00$\pm$0.21 given the fact that neither  bump nor dip is observed in $p \bar{p}$ data within uncertainties. It leads to a more than 3$\sigma$ difference between $pp$ and $p \bar{p}$ elastic data (assuming a flat behavior above $\sqrt{s} = 100 GeV$).

\begin{figure}
\begin{center}
\epsfig{file=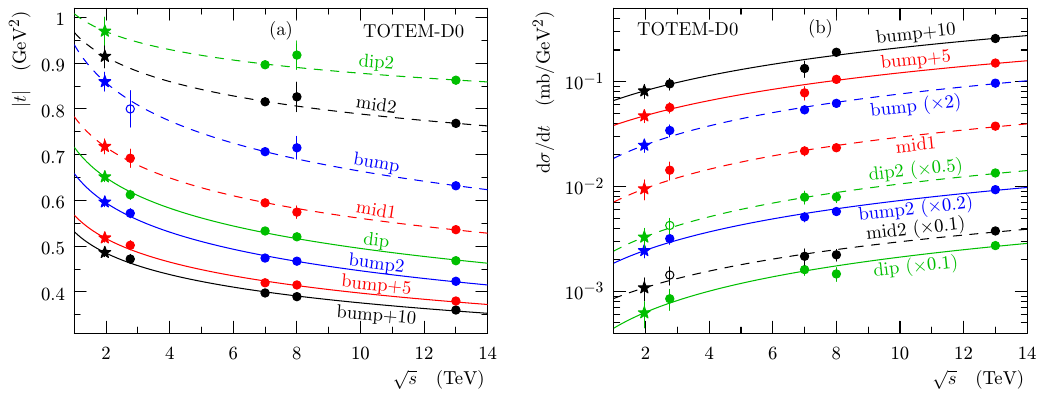,width=13.5cm}
\caption{Variation of $t$ and $d\sigma/dt$ values for reference points at $\sqrt{s}=$2.76, 7, 8 and 13 TeV extrapolated down to Tevatron energies $\sqrt{s}=1.96$ TeV.}
\label{variation}
\end{center}
\end{figure}

An even more quantitative comparison is possible using the eight characteristic points described above. The values of $|t|$ and $d\sigma/dt$ for these eight points are shown in Fig.~\ref{variation} as a function of $\sqrt{s}$ of 2.76, 7, 8, and 13 TeV extrapolated down to the Tevatron $\sqrt{s}=$1.96 TeV, using a two parameter fit
\begin{eqnarray}
|t| &=& a \log (\sqrt{s}{\rm [TeV]}) + b  \label{eqn2} \\
(d\sigma/dt) &=& c \sqrt{s}~{\rm [TeV]} + d \label{eqn3}.
\end{eqnarray}
Note that the extrapolation at 1.96 TeV using a variety of different fitting functions is well within uncertainties. 

\begin{figure}
\begin{center}
\epsfig{file=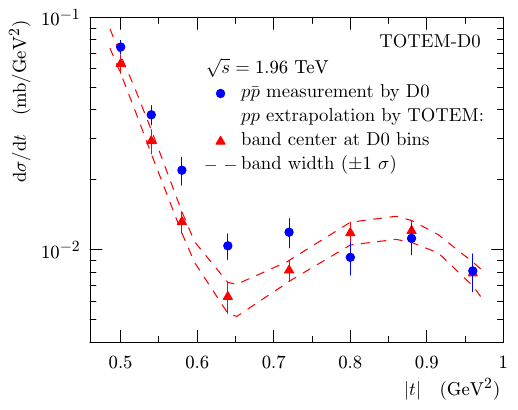,width=9.5cm}
\caption{Comparison between the elastic $d \sigma/dt$ cross section measurements at $\sqrt{s}=1.96$ TeV by the D0 collaboration for $p \bar{p}$ interactions and by the TOTEM collaboration for $pp$ interactions, extrapolated from measurements at 2.76, 7, 8 and 13 TeV. The difference can be explained by the existence of the odderon.}
\label{odderon}
\end{center}
\end{figure}

Using the extrapolated characteristic points from the TOTEM measurements, it is now possible to predict the elastic $pp$ cross sections at Tevatron energies in the same $|t|$ bins as for the D0 measurement. Differences in normalization are taken into account by
adjusting TOTEM and D0 data sets to have the same cross sections at the optical point (OP)
$d\sigma/dt(t = 0)$ (OP cross sections are expected
to be equal if there are only $C$-even exchanges). We first 
predict the $pp$ total  cross section from an extrapolated fit to TOTEM data with a $\chi^2=$0.27
\begin{eqnarray}
\sigma_{tot} = a_2 \log^2 \sqrt{s}{\rm [TeV]} +b_2, 
\end{eqnarray}
that leads to an estimate of $pp$ $\sigma_{tot}=$82.7 $\pm$ 3.1 mb at 1.96 TeV.
Different parametrizations (for instance a 3-parameter formula including a $\log s$ term) lead to similar results within uncertainties.  From the extrapolated TOTEM $pp$ $\sigma_{tot}$ at 1.96 TeV, it is possible to extract $d\sigma/dt(t=0)$ using
\begin{eqnarray}
\sigma_{tot} = \sqrt{\frac{16 \pi (\hbar c)^2}{1+\rho^2}  \left( \frac{d \sigma}{dt} \right)_{t=0}}. \label{rhoformula}
\end{eqnarray}
Assuming the ratio of the imaginary and the real part
of the elastic amplitude $\rho=$0.145, as taken from COMPETE~\cite{compete}, 
this leads to $d\sigma/dt(t=0)$ at the OP of 357.1 $ \pm$ 26.4 mb/GeV$^2$ from the TOTEM data (since the dependence on $\rho$ is squared in Eq.~\ref{rhoformula}, the result does not depend strongly on the exact value of $\rho$). 
The D0 collaboration measured the optical point of $d \sigma/dt$ at small $t$ to be 341$\pm$48 mb/GeV$^2$, and we thus rescale the TOTEM data by the ratio 0.954 $\pm$ 0.071, which is compatible with 1.0 within uncertainties.
Please note that we do
not claim that we performed a measurement of $d\sigma/dt$ at
the OP at $t = 0$ (it would require additional measurements  closer to $t = 0$), but
we use the two extrapolations in order to obtain
a common and somewhat arbitrary normalization point. 

The comparison between the D0 $p \bar{p}$ elastic cross sections measurements and the extrapolated TOTEM $pp$ elastic data down to Tevatron energies is shown in Fig.~\ref{odderon}.  The $\chi^2$ test with six degrees of freedom yields a $p$-value of 0.00061, corresponding to a significance of 3.4$\sigma$. This result can be 
combined with the independent evidence of the 
odderon found by TOTEM using the  $\rho$ and total cross section measurements at low $t$~\cite{rho} which leads to a 5.3 to  5.7$\sigma$ discovery~\cite{ourpaper} depending on the models~\cite{compete,models} for $\rho$.

\section{A new scaling in elastic data}

In this section, we will discuss a new scaling that appears in elastic $d \sigma/dt$ cross sections~\cite{scaling1,scaling2}. 
The fact that we can use the same fitting formulae (see Eq.~\ref{eqn2} and \ref{eqn3}) to describe the evolution of all characteristic points as a function of $\sqrt{s}$ shows that all $d\sigma/dt$ cross sections for $\sqrt{s}=$2.76, 7, 8 and 13 TeV are stackable between each other as suggested in Fig.~\ref{totem}.

\begin{figure}
\begin{center}
\epsfig{file=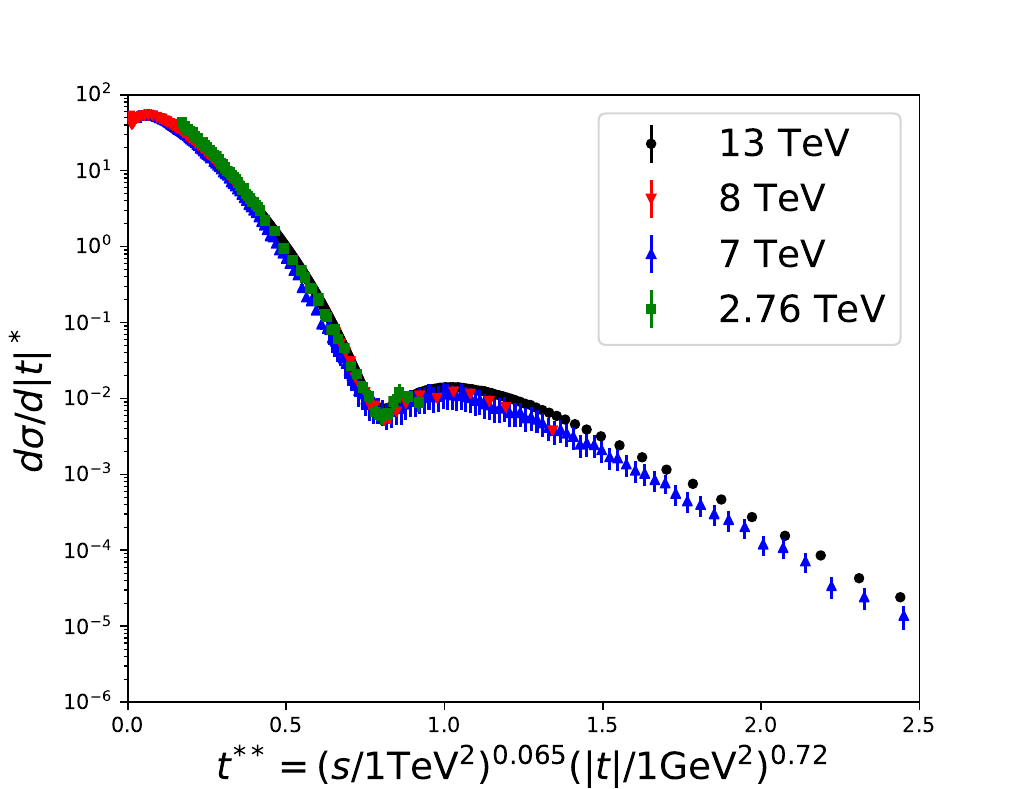,width=6.2cm}
\epsfig{file=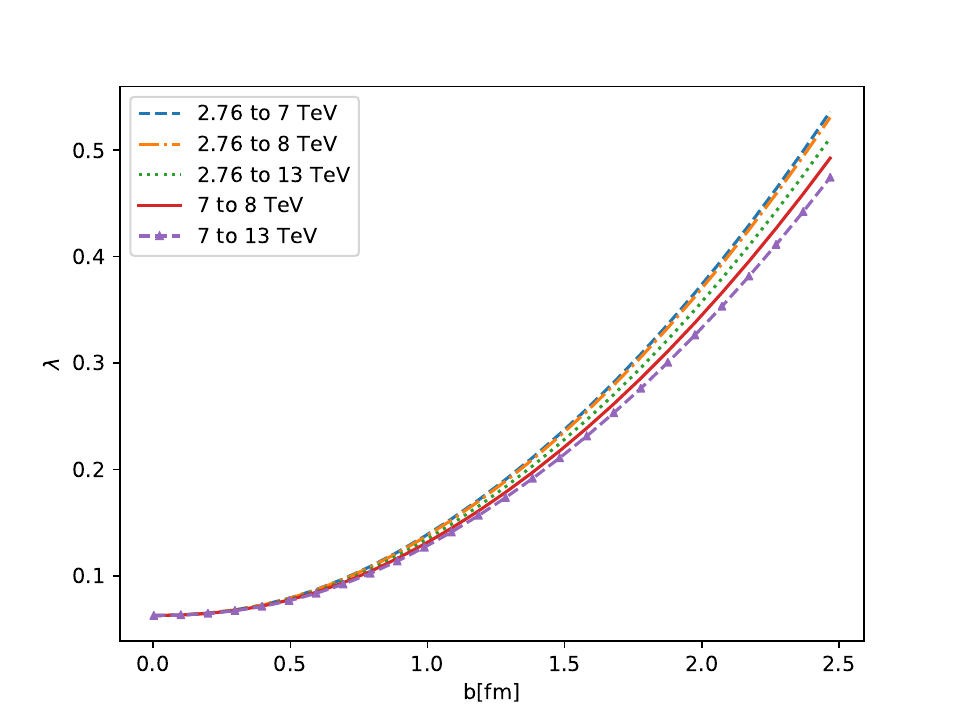,width=6.2cm}
\caption{Left: $d \sigma/dt^*$ as a function of $t^{**}=t^* / s^{0.215}$ where $t^*=(s/|t|)^{0.28} \times |t|$ showing a scaling in elastic $d \sigma/dt$ cross sections measured by the TOTEM collaboration at 2.76, 7, 8 and 13 TeV. Right: Power growth exponent $\lambda$ as a function of $b$ for various reference $\sqrt{s}$ pairs.}
\label{scaling}
\end{center}
\end{figure}

In order to introduce a new scaling in elastic data, we first introduce the variable $t^*=(s/|t|)^A \times |t|$, inspired by geometric scaling in terms of saturation models~\cite{scalinggeneral}, and 
$t^{**}=t^* / s^B$, $A$ and $B$ being parameters to be fitted to data~\cite{scaling1}.  In fact, $A$ and $B$ are correlated, and we have a full valley of parameters leading to similar scalings ($B=A-0.065$), that leads to one single parameter fit to data ($A=0.28$). As shown in Fig.~\ref{scaling}, $d \sigma/dt^*$ shows scaling as a function of $t^{**}$. The $s$ dependence on $d \sigma /dt$ is imposed by scaling as $s^{-\alpha}$ where $\alpha =-A(A-1.065)/ (1-A)=0.305$. Scaling is not supposed to work perfectly
at low $|t|$ that corresponds to the QED Coulomb and non-perturbative QCD region, and at high $|t|$ in the perturbative QCD
domain.

The next step is to compute the profile function $\Gamma$ in the impact parameter space. We first perform a fit of the elastic amplitude $A$ using the definition
\begin{eqnarray}
\frac{\mathrm{d}\sigma}{\mathrm{d}|t|}= \frac{1}{16 \pi s^2 } |A(s,t)|^2  = |\mathcal{A}(s,t)|^2.
\end{eqnarray}
The fit formula is the following
\begin{eqnarray}
\mathcal{A}(s,t) &=& i\big(\mathcal{A}_1 (s,t) + \mathcal{A}_2(s,t) \big)e^{i\theta}
\\
\mathcal{A}_{1}(s,t) &=& N_1 (s) e^{-B_1(s)|t|}  \\ \mathcal{A}_{2}(s,t) &=& N_2 (s) e^{-B_2(s)|t|}e^{i\phi} 
\end{eqnarray}
where $N_1(s) = N_1^0 s^{\alpha/2}$, $N_2(s) = N_2^0 s^{\alpha/2}$, $B_1(s) = B_1^0 s^{\gamma/2}$ and $B_2(s) = B_2^0 s^{\gamma/2}$. There are six free parameters in the fit, namely $N_1^0$, $N_2^0$, $B_1^0$, $B_2^0$, $\phi$, and $\theta$. $\alpha = 0.305 $ and $\gamma/2 \equiv  0.065/(1-A) = 0.065/0.72 \approx 0.09$ being  fixed by scaling.
The fit quality is quite good with a $\chi^2/dof=$1.08 for $0.2<t^{**}<1.5$ in the dip-bump region for 476 data points.
We then compute the profile function $\Gamma$ in the impact $b$-parameter space using the following equation \begin{eqnarray}
\mathrm{Re}(\Gamma(s,b)) = \frac{1}{4 \pi i s} \int_0^{\infty} dq \, q  \, J_0(qb) \,  A(s,t=-q^2) \label{profile}
\end{eqnarray}
where $J_0$ is the 0th order Bessel function.
 
 We define $\lambda$ as a function of the ratio of two values of $\Gamma$ for two values of $\sqrt{s}$ as
\begin{eqnarray}
 \lambda = \frac{1}{\ln (s_1/s_2)} \ln \Big(\frac{\mathrm{Re} \Gamma(s_1,b)}{\mathrm{Re} \Gamma(s_2,b)}\Big). 
\end{eqnarray}
The values of $\lambda$ as a function of $b$ for different $\sqrt{s}$ are shown in Fig.~\ref{scaling}, right.
$\lambda = (\alpha - \gamma)/2=0.06$ when $b \rightarrow 0$ as predicted by scaling. This means that scaling predicts a universal behavior of $\lambda$ at small $b$.
Values of $\lambda$ at small $b$ are compatible with expectations from a dense object, such as a black disc,  and reach higher values around 0.3 for $b = 1$ fm, which recalls the power-law exponent in the small-$x$ limit of QCD, described by the perturbative Balitsky Fadin Kuraev Lipatov (BFKL) evolution equation~\cite{bfkl} at next-to-leading logarithmic accuracy.
In this sense, we can interpret our results as the presence of dense gluonic objects in the proton at high energy. The density of these objects in the proton can be small, but the density of the gluons inside  can be large.

An additional scaling was described in Ref.~\cite{scaling2}.  The bump over dip elastic cross section $d\sigma/dt$ ratio is constant at high energies and the position ratio in $|t|$ between the bump and the dip is also constant between ISR and LHC energies.   This leads to a new scaling of $W^{-\alpha} \frac{d\sigma_{el}}{dt} (\tau)$ as a function of $\tau = W^{\beta} |t|$ with $\alpha=0.66$ (with large uncertianties) and $\beta = 0.1686$.

\section{High gluon density regime at the LHC}

\begin{figure}
\begin{center}
\epsfig{file=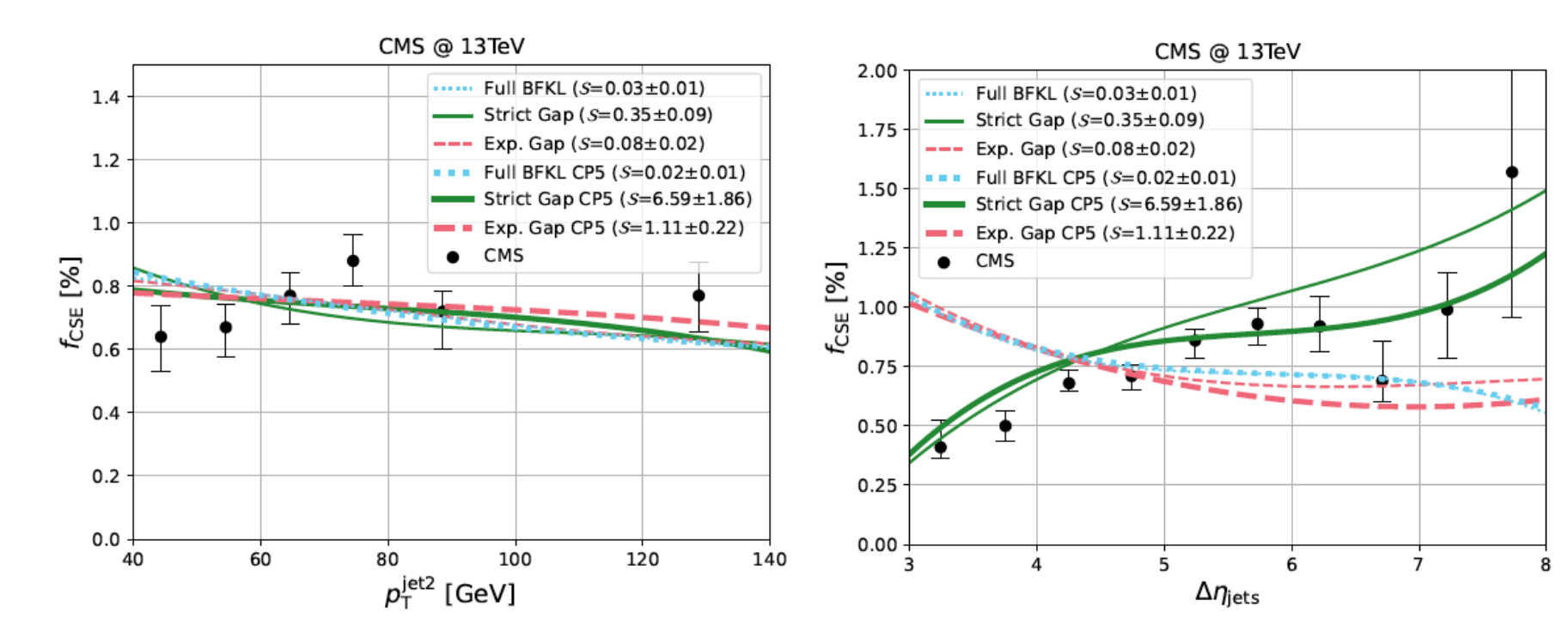,width=11.0cm}
\epsfig{file=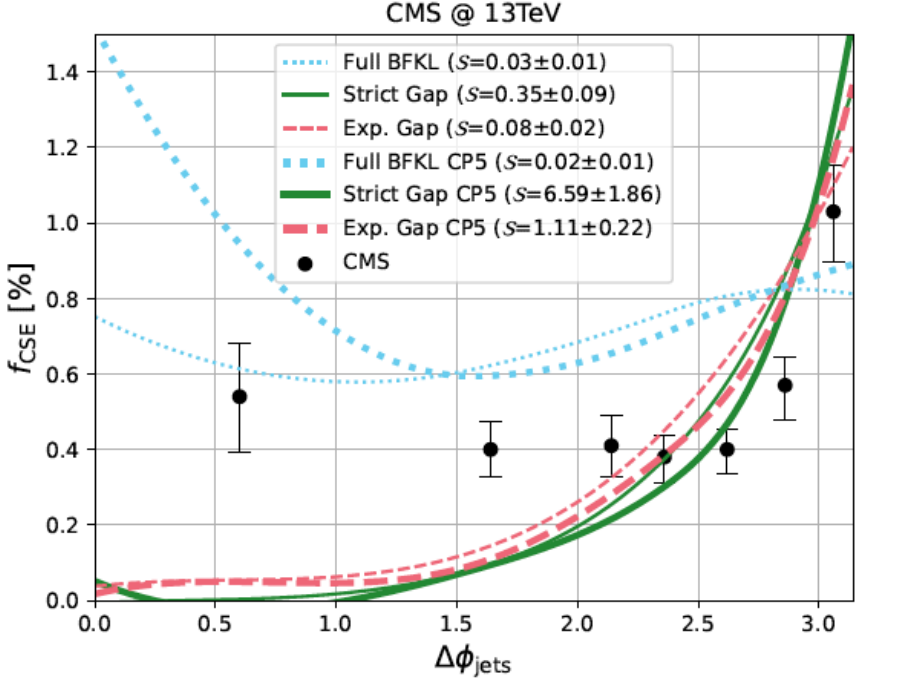,width=5.5cm}
\caption{Fraction of gap between jet events as a function of the transverse momentum of the second leading jet, the interval in rapidity and in azimuthal angle between the two jets, compared to different models.}
\label{jetgapjet}
\end{center}
\end{figure}

\begin{figure}
\begin{center}
\epsfig{file=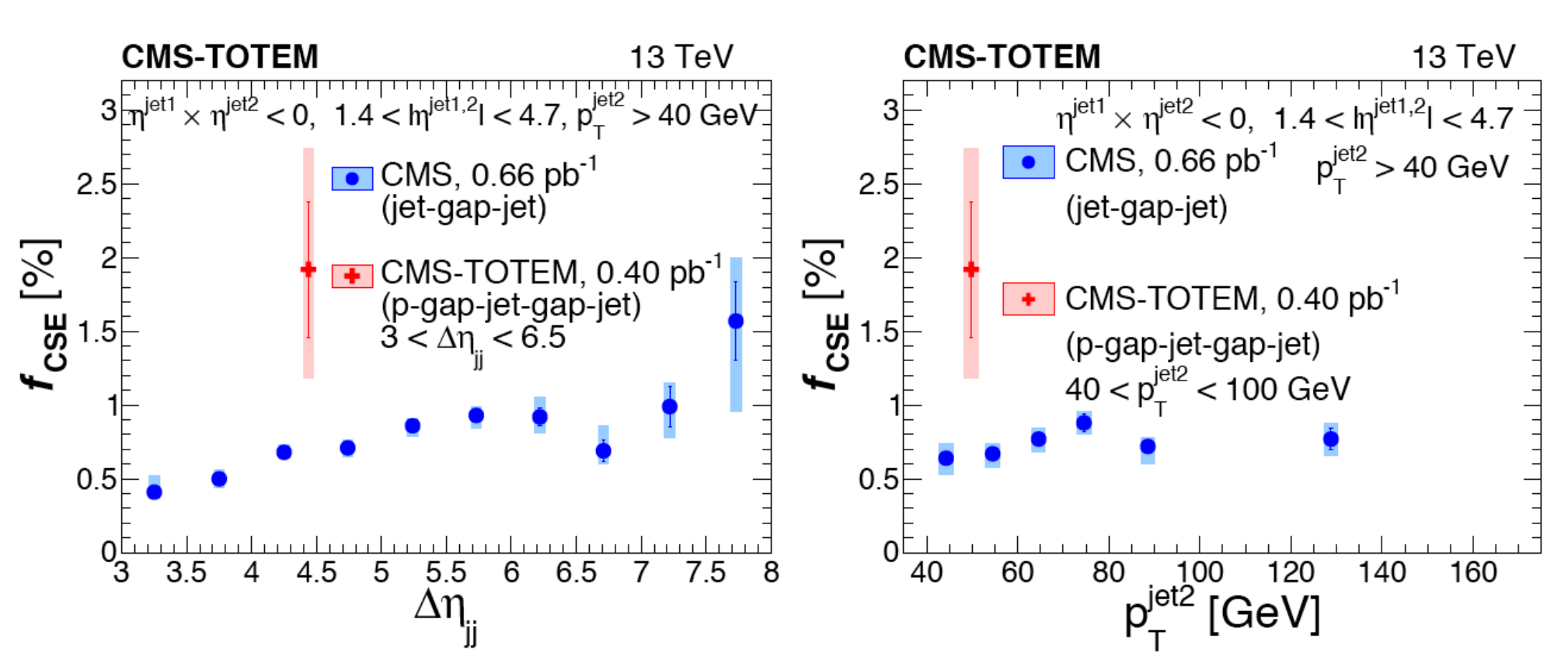,width=10.5cm}
\caption{Fraction of gap between jet events as a function of the interval in rapidity between the two jets and the transverse momentum of the second leading jet for inclusive (blue points) and diffractive (red points) events when a proton is detected in the TOTEM roman pots. Eleven events were observed with 0.7 pb$^{-1}$.}
\label{jetgapjetdiff}
\end{center}
\end{figure}

In this section, we describe new observables that are sensitive to the high gluon density regime at the LHC. We consider two evolution equations to describe the structure of protons or heavy ions, namely the Dokshitzer Gribov Lipatov Altarelli Parisi (DGLAP)~\cite{dglap} evolution in resolution $Q^2$ (resumming terms in $\alpha_S \log Q^2$, resolving ``smaller" partons at high $Q$) and the BFKL~\cite{bfkl}  evolution in energy $x$ (resumming terms in $\alpha_S \log 1/x$  leading to large parton densities at small $x$). Ultimately, at very small $x$, the domain of saturation or gluon recombination can be reached, that leads to slower evolution of gluon densities as a function of energy.

Many observables have been suggested to look for the high gluon density regime at small $x$ such as the measurement of large interval in rapidity between jets, the so-called Mueller Navelet jets~\cite{forward,mnjets}, the measurement of the azimuthal angle between jets~\cite{soeren}, or the measurement of gap between jets~\cite{jgj1,jgj2}. We will focus on this last measurement performed by the CMS collaboration, namely the cross section of two jets separated by a large interval in rapidity where the region ($-1<\eta<1$) is devoid of any charged particle production (requesting the absence of reconstructed tracks with a transverse momentum above 200 MeV)~\cite{jgj1}. Requesting a gap between the two jets is only possible using special runs at the LHC with small pile up.
The cross section predicted by NLO DGLAP is expected to be negligible for large gaps, below what can be measured, and is thus a clean test of the BFKL resummation. In Fig.~\ref{jetgapjet}, we display the fraction of gap between jet events as a function of the transverse momentum of the second leading jet, the difference in rapidity and in azimuthal angle between the two jets. Data are compared to different models~\cite{jgj2}. We will only discuss two of them, namely the experimental gap (in dashed red line) and the strict gap definition (in dark green full line), assuming the BFKL evolution equation~\cite{bfkl}. The first model corresponds to the experimental definition of the gap as performed by the CMS collaboration, namely the absence of charged particles above 200 MeV in the gap region ($-1 < \eta < 1$). It leads to a bad description of data especially for the difference in rapidity between the two jets. 
The second model corresponding to a strict gap definition (no particle above 1 MeV in the gap region) leads to a good description of data except at very low $\Delta \phi$. It is of course not possible to achieve this threshold experimentally. However, to understand why it leads to a fair description of data, we noticed that the distribution of charged particles from PYTHIA8~\cite{pythia8} in the gap region ($-1 < \eta < 1$) with initial state radiation is very large. Particles emitted at large angle with $p_T > 200$ MeV from initial state radiation have large influence on the gap presence or not, and thus on the gap definition (experimental or strict). Retuning the amount of initial state radiation allows obtaining a good agreement for the ratio of gap between jets events using the experimental gap definition~\cite{jgj2}.

We also computed a full calculation of jet gap jet cross section at next-to-leading logarithm accuracy using the next-to-leading orders BFKL kernel and impact factors~\cite{jgj3}. Compared to the results obtained with leading order impact factors, the effect is quite small leading to a higher cross section by about 20\% at high jet $p_T$ and shows even a smaller effect on the rapidity dependence. The total uncertainties due to the scale variation are also smaller, of the order of 15-20\%.

The CMS collaboration measured the subsample of gap between jets events requesting in addition at least one intact proton on either side of CMS in the TOTEM roman pot detectors~\cite{jgj1}. Eleven events were observed with a gap between jets of ($-1< \eta < 1$) and at least one proton tagged with $\sim 0.7$ pb$^{-1}$.  
Jet gap jet events were observed for the first time by the CMS collaboration. This leads to very clean events for jet gap jets measurements since the effects of multi-parton interactions  are suppressed. This might be the ideal way to probe low $x$ (BFKL) resummation effects~\cite{jgjdiff} and would benefit from more statistics  (about 10 pb$^{-1}$ for single diffractive and 100 pb$^{-1}$ for double pomeron exchange events of low pile up data would be needed to perform some measurements of differential distributions).

Using the measurement of gap between jets at the LHC, we now have some evidence of BFKL resummation of the high gluon density regime at the LHC and we can study if we can reach the domain of saturation.

\section{Possible observation of saturation phenomenon in $Pb Pb$ collisions at the LHC}

In this section, we will discuss the possible observation of saturation phenomenon at the LHC. When the value of $x$ becomes small, the number of gluons in the proton (or heavy ions) becomes very large. The usual linear equations of QCD (DGLAP or BFKL) are not longer valid and one needs to consider recombination of gluons leading to non-linear evolution equations such as the Balitsky-Kovchegov (BK)~\cite{bk} equation. This leads to saturation phenomenon or slower evolution of the energy dependence of the gluon density. The idea is to look for the best observables sensitive to saturation in PbPb interactions.

\subsection{Exclusive vector meson production}

\begin{figure}[t!]
    \centering
        \includegraphics[width=0.49\textwidth]{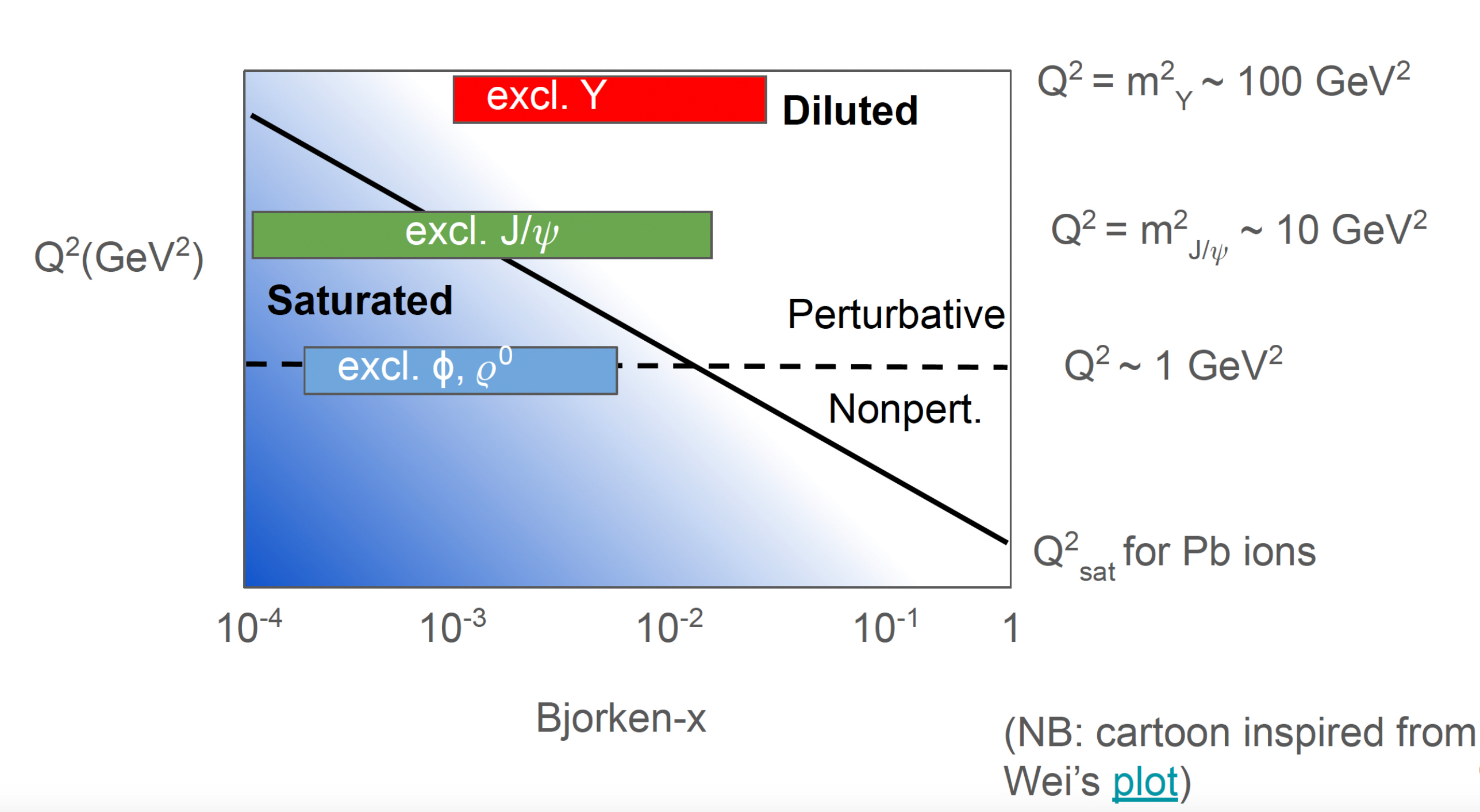}
    \includegraphics[width=0.49\textwidth]{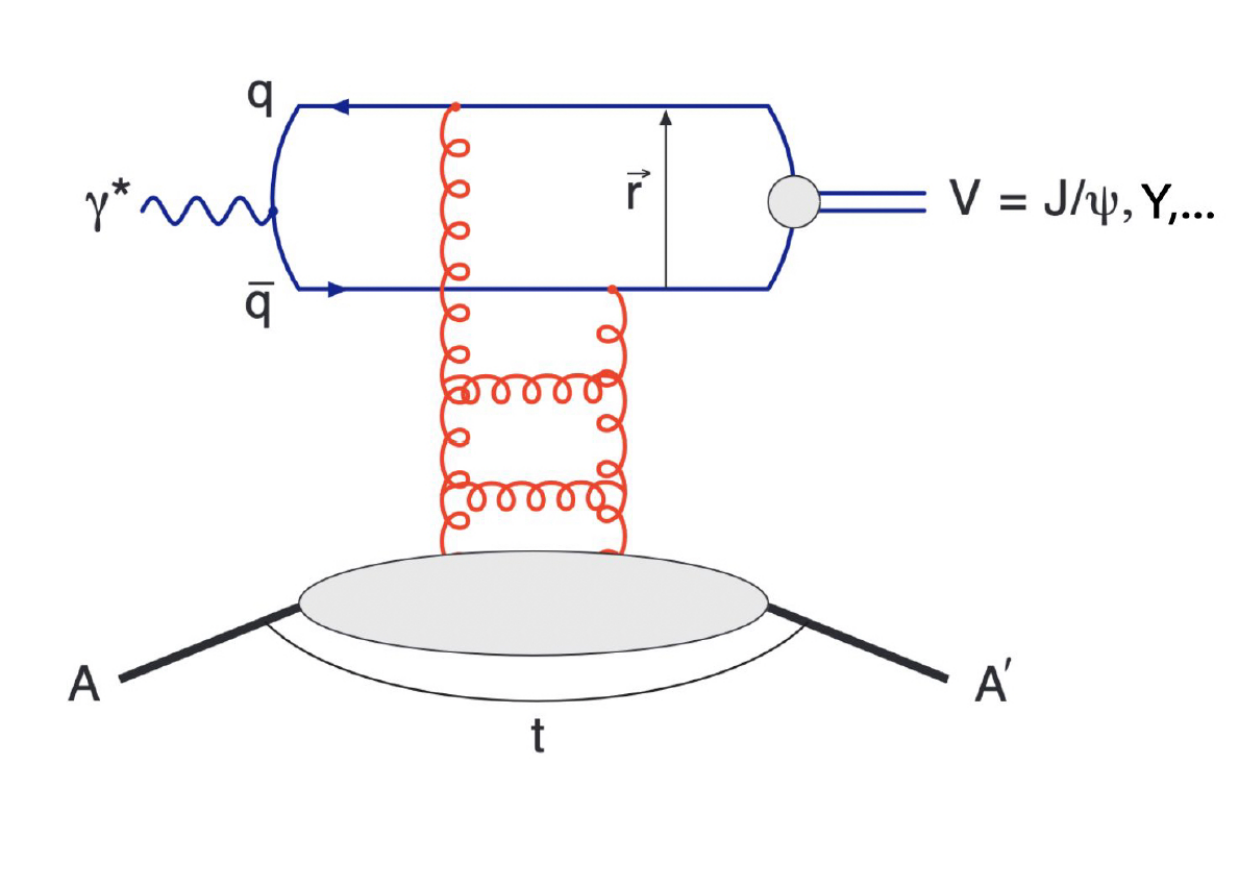}
    \caption{Left: Scheme of sensitivity to saturation for vector meson production. Right: Exclusive vector meson production in $\gamma$ Pb interactions.
        }
    \label{vectormeson}
\end{figure}

\begin{figure}[t!]
    \centering
    \includegraphics[width=0.32\textwidth]{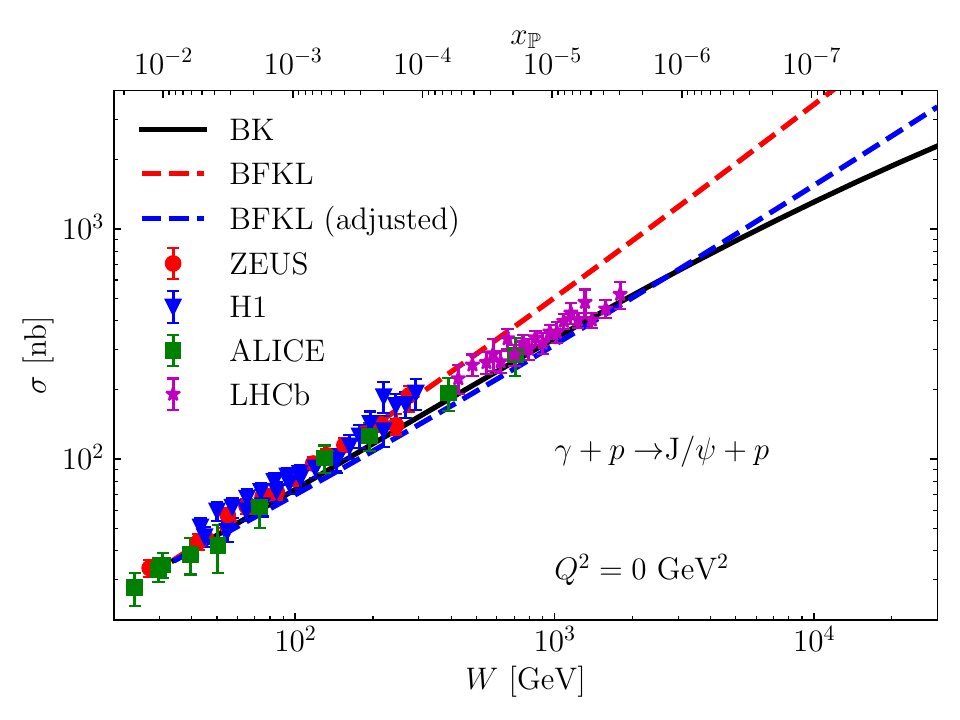}
    \includegraphics[width=0.32\textwidth]{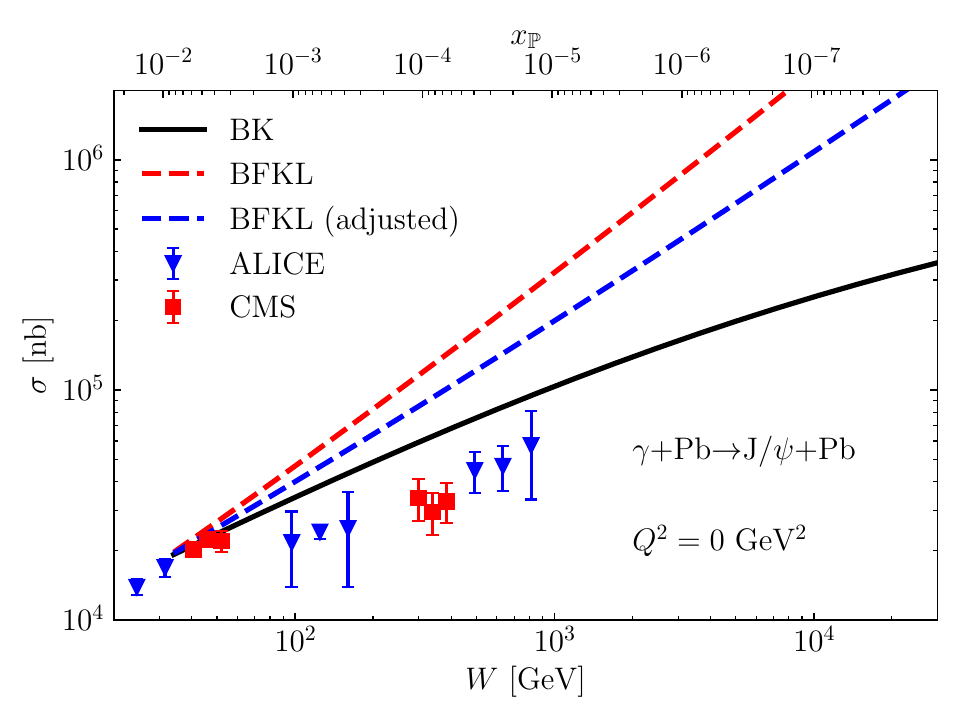}
    \includegraphics[width=0.32\textwidth]{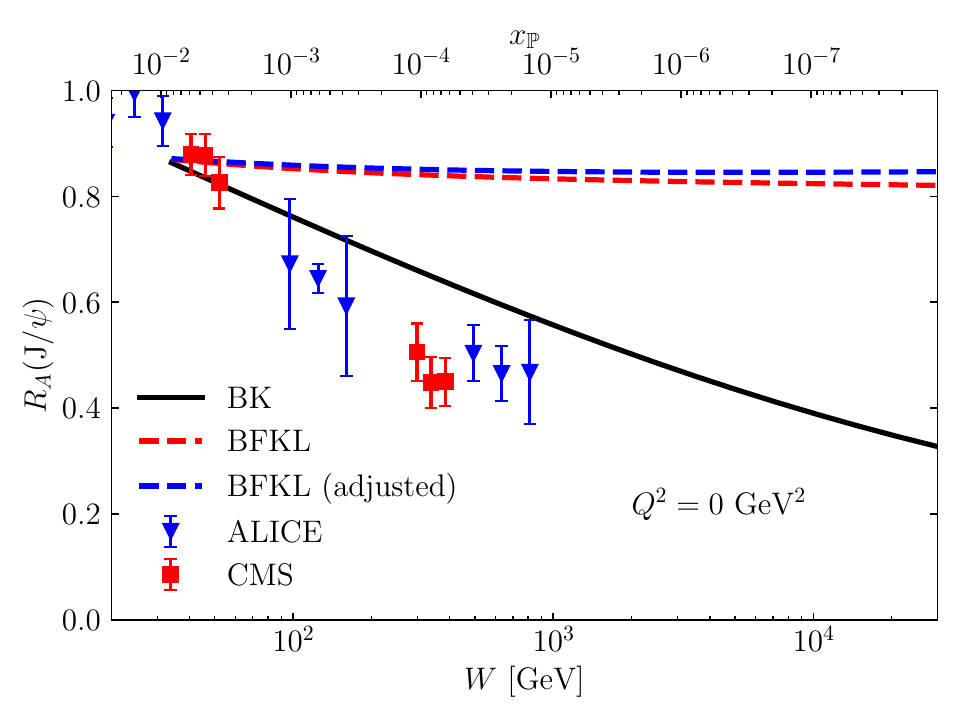}
    \caption{Exclusive $J/\Psi$ production cross section as a function of the center-of-mass energy $W$. 
    Left: $\gamma$p, 
    Center: $\gamma$Pb,
  Right: Nuclear suppression factor. Predictions without (BFKL) and with saturation (BK) are compared with measurements at HERA (H1 and ZEUS) and at the LHC (ALICE, CMS and LHCb).
        }
    \label{jpsi}
\end{figure}

A scheme of the sensitivity to saturation for vector meson production is shown in Fig.~\ref{vectormeson}, left. In order to see saturation effects, one needs to have a dense object with a high gluon content, such as Pb. In addition, we need a characteristic scale (for instance the mass of the vector meson) which is above the perturbative scale (of the order of 1 GeV) to trust perturbative calculation and below the saturation scale $Q_S$. $Q_S$ is of the order of 1 GeV for protons
 whereas it is increased by $A^{1/3}$ for heavy ions ($Q_S \sim  5.8$ GeV for Pb). Exclusive $J/\Psi$ production with a scale squared $m_{J/\Psi}^2 \sim 10$ GeV$^2$ is ideal to observe saturation being below $Q_S^2$. On the contrary, the exclusive production of $\Upsilon$ meson should not be showing much saturation effects since its scale $M_{\Upsilon}^2 \sim 100$ GeV$^2$ is above $Q_S^2$. The exclusive production of $\rho$ mesons has a scale $m_{\rho}^2 \sim 1$ GeV$^2$ well below
  $Q_S^2$, but of the same order of magnitude as the lower limit of the perturbative scale, which makes it difficult to trust perturbative calculation. The diagram showing vector meson production in $\gamma A$ interactions (the $\gamma$ being emitted by the Pb ion) for diffractive events where $A$ is intact after the collision is shown in Fig.~\ref{vectormeson}, right. In the case  of inclusive production, the two gluon exchange should be replaced by a single one. Using the vector meson production,  
one probes the gluon content of $A$ which could be a proton (resp. Pb)  in the case of pPb (resp. PbPb) interactions. Saturation effects are expected for instance in $J/\Psi$ exclusive production when one probes the gluon content in Pb 
($m_{J/\Psi}^2 < Q_S^2\sim 34$ GeV$^2$) but not for protons ($m_{J/\Psi}^2 > Q_S^2=1$ GeV$^2$). We have a similar situation when we consider the exclusive production of $c \bar{c}$ and $b \bar{b}$ where the typical scale can be below $Q_S$ in $\gamma$Pb interactions.

To compute the exclusive production of $J/\Psi$ or $\Upsilon$,  the process can be factorized in two parts, namely
the photon fluctuation into a quark-antiquark pair, and the quark-antiquark pair interacting with the target hadron.
The first part is described by the light-cone wave function of the photon computed using  QED and 
the second part of the process is given by the dipole scattering amplitude which is 
non-perturbative. 
However, its energy dependence can be described using perturbative evolution equation with and without saturation effects using the BK~\cite{bk} or the  BFKL~\cite{bfkl} equations.
The dipole amplitude at the starting scale of the evolution is fixed by fitting the predictions to measured data for instance the cross section of vector meson production at HERA~\cite{jani}. We also take into account $b$ impact parameter dependence in the dipole amplitude. We
modify the traditional Mc Lerran Venugopalan model to take into the $b$ dependence with a gaussian dependence of the thickness function~\cite{salazar}
\begin{eqnarray}
T(b)= \frac{1}{2 \pi B_P} e^{-b^2/(2B_P)} 
\end{eqnarray}
There is a large difference between taking into account or not the $b$ impact parameter dependence for $\gamma$p cross sections and the effect is smaller for $\gamma $Pb interactions since a nucleus is much larger than a proton and neglecting impact parameter dependence is more justified~\cite{salazar}. Technically, we will study the effect of saturation in the following by comparing the evolution of cross sections as a function of energy $W$ using either the BK or the BFKL equations (which is equivalent of neglecting the gluon recombination term in BK) including running $\alpha_S$.

In Fig.~\ref{jpsi}, we display the exclusive $J/\Psi$ production cross section as a function of the center-of-mass energy $W$ for $\gamma$p (left), $\gamma$Pb (center) interactions and the nuclear suppression factor (right) from the H1~\cite{h1}  and ZEUS~\cite{zeus} collaborations at HERA and the ALICE~\cite{alice}, CMS~\cite{cms} and LHCb~\cite{lhcb} collaborations at the LHC, compared to predictions with and without saturation~\cite{jani}. For $J/\Psi$ production in $\gamma$p interactions, small differences are observed between the BK and BFKL evolution and there is no sign for saturation as expected since one probes the gluon in the proton. On the contrary, $\gamma$Pb cross sections clearly favor saturation models. The difference between the BFKL and BFKL-adjusted predictions correspond to a small variation in $\Lambda_{QCD}$ to obtain a better description of data in $\gamma$p interactions. It seems that data even show more saturation effects than our calculation but it is worth noting that an impact parameter NLO BK equation would be needed in order to compute more precise predictions for $\gamma$Pb interactions. Our goal was more to show that the linear BFKL evolution equation seems to be disfavored and some additional effect such as saturation is needed to describe data. However, alternative approaches such as nuclear PDFs including shadowing can also lead to a good description of $J/\Psi$ exclusive cross sections, and we will propose an additional measurement to distinguish between both models in the following subsection.

\begin{figure}[t!]
    \centering
    \includegraphics[width=0.32\textwidth]{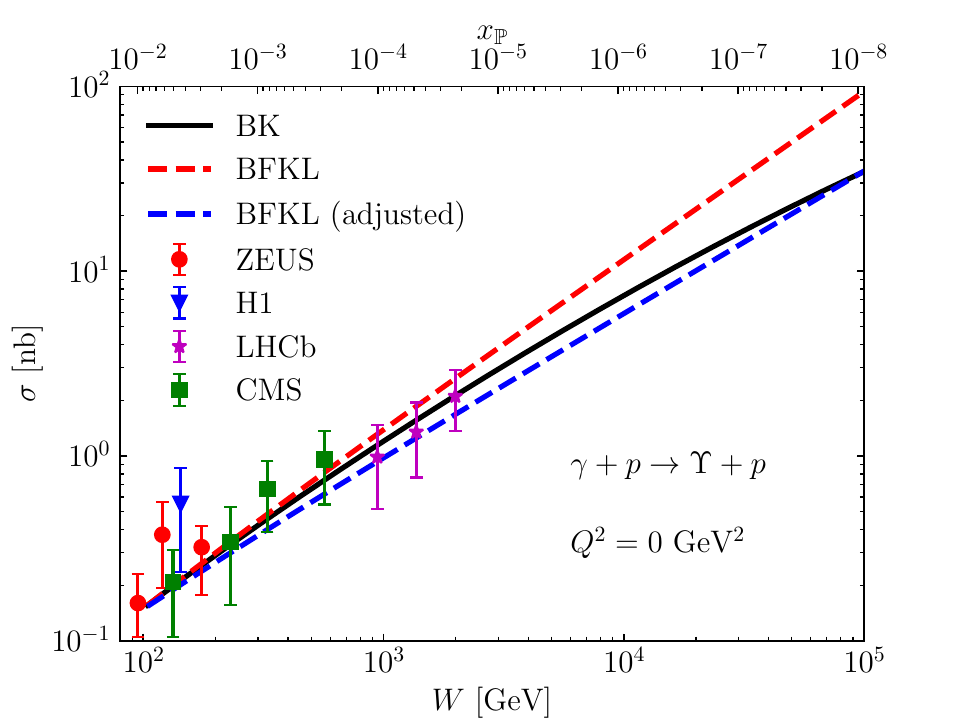}
    \includegraphics[width=0.32\textwidth]{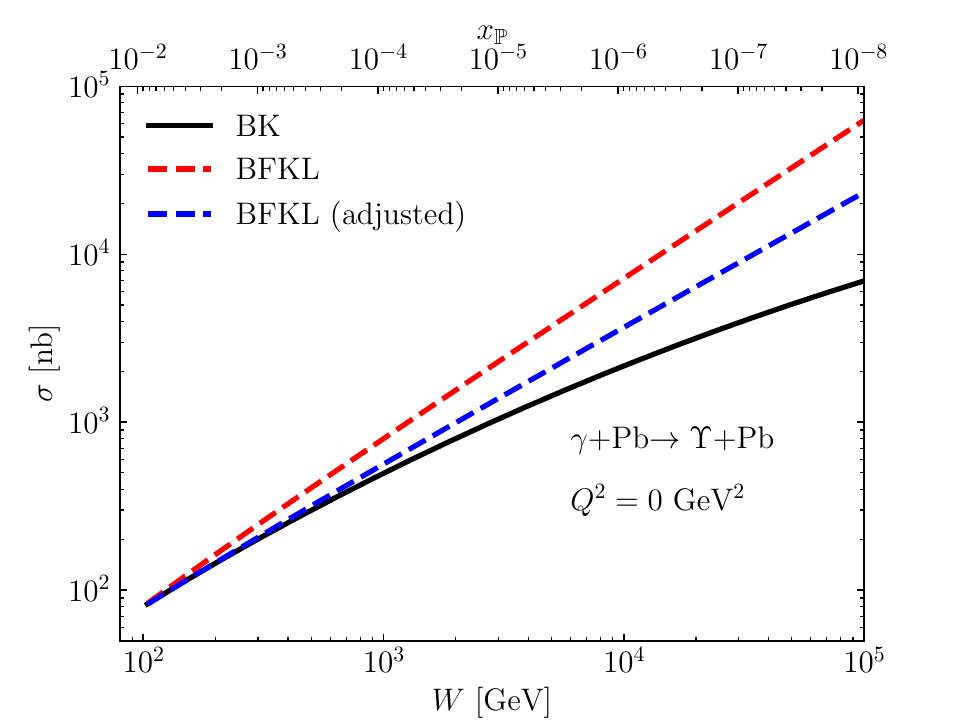}
    \includegraphics[width=0.32\textwidth]{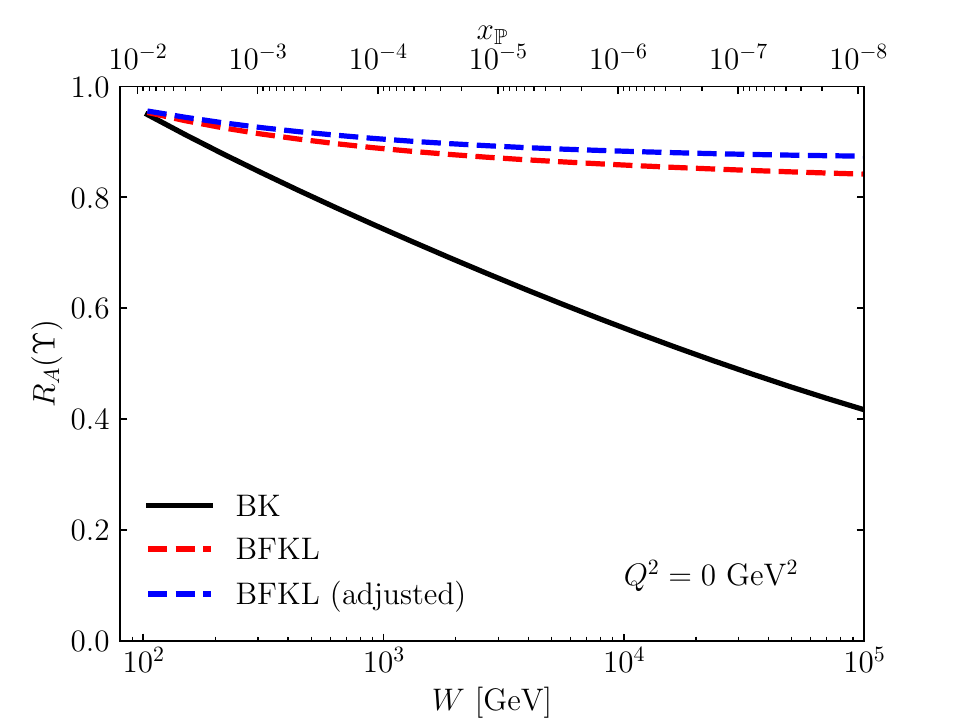}
    \caption{Exclusive $\Upsilon$ production cross section as a function of the center-of-mass energy $W$. 
    Left: $\gamma$p, 
    Center: $\gamma$Pb,
    Right: Nuclear suppression factor. Predictions without (BFKL) and with saturation (BK) are compared with measurements at HERA (H1 and ZEUS) and at the LHC (CMS and LHCb}
    \label{upsilon}
\end{figure}

In Fig.~\ref{upsilon}, we display similar results for exclusive $\Upsilon$ production as a function of $W$ for $\gamma$p (left), $\gamma$Pb (center) interactions and the nuclear suppression factor (right). As expected, there is a good agreement between the BK and BFKL predictions and data for $\gamma$p cross sections and small differences are observed between both approaches. In $\gamma$Pb interactions, small differences are also predicted between the BK and BFKL evolutions because of the $\Upsilon$ vector meson mass for $W\sim$1 TeV. Recent measurements from the CMS collaboration are compatible with this prediction~\cite{upsiloncms}.

\subsection{$c \bar{c}$ and $b \bar{b}$ production}

\begin{figure}[t!]
    \centering
    \includegraphics[width=0.95\textwidth]{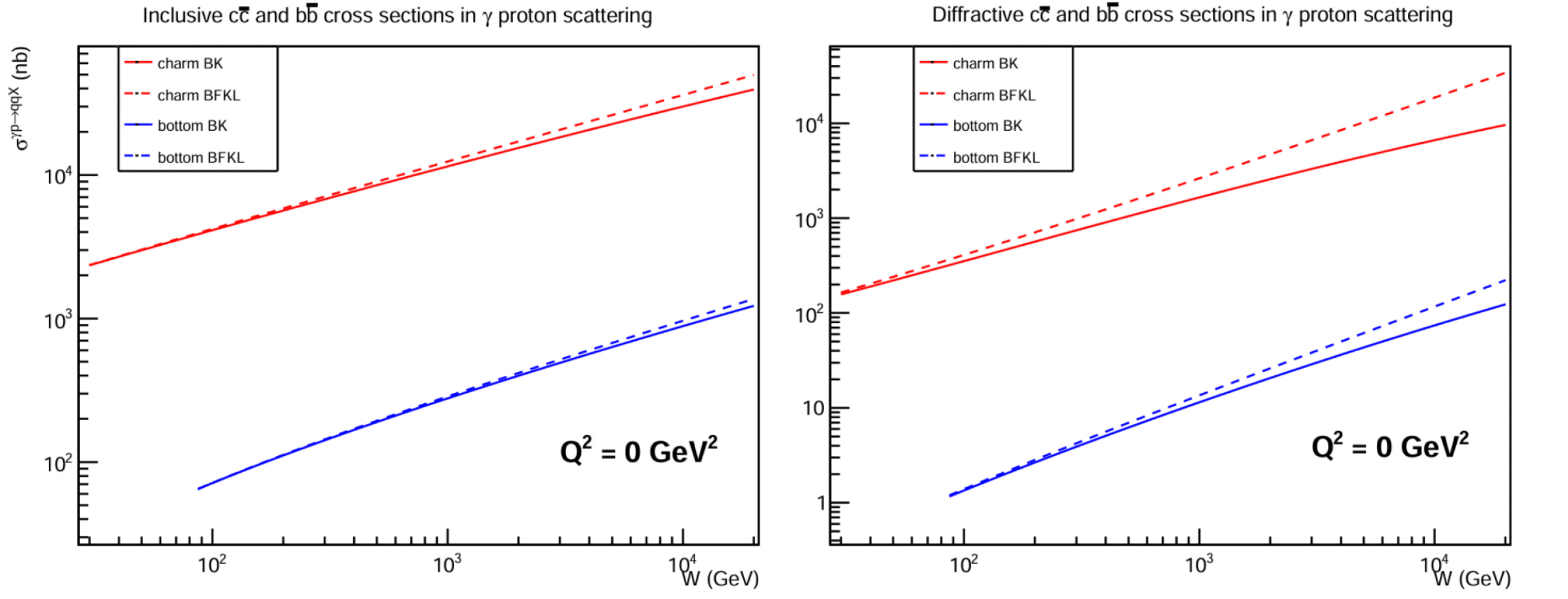}
    \caption{Inclusive (left) and diffractive (right) $c \bar{c}$ (in red) and $b \bar{b}$ (in blue) production cross sections for $\gamma p$ interactions. The BFKL predictions are in dashed lines and the BK ones including saturation effects in full line.
        }
    \label{ccbargammap}
\end{figure}

\begin{figure}[t!]
    \centering
    \includegraphics[width=0.95\textwidth]{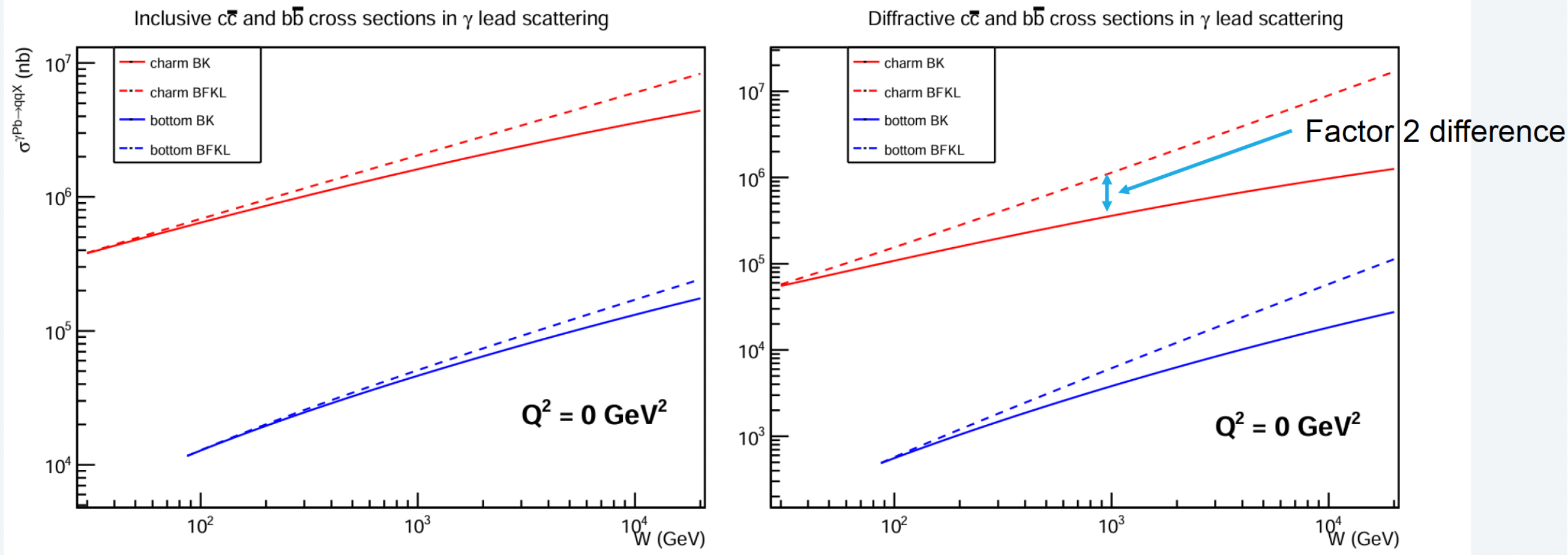}
    \caption{Inclusive (left) and diffractive (right) $c \bar{c}$ (in red) and $b \bar{b}$ (in blue) production cross sections for $\gamma Pb$ interactions. The BFKL predictions are in dashed lines and the BK ones including saturation effects in full line.
        }
    \label{ccbargammalead}
\end{figure}

In this subsection, we discuss additional possible measurements that are sensitive to saturation effects and that might allow to distinguish between color glass condensate (saturation) and nuclear PDFs and shadowing approaches~\cite{jarno}. We consider the production of $c \bar{c}$ (in red) and $b \bar{b}$ (in blue) in $\gamma$p and $\gamma$Pb interactions respectively in Fig.~\ref{ccbargammap} and \ref{ccbargammalead} using the BFKL (dashed line) and the BK (full line) evolution. The left figure correspond to the inclusive cross section (when one gluon is exchanged with the proton or with Pb) whereas the right one to the diffractive one when the proton or $Pb$ is intact after the interaction. As expected, not much difference is observed for $\gamma$p interactions between the BFKL and BK evolution whereas a difference of about a factor two is observed for $\gamma$Pb interactions for diffractive production at $W\sim$1 TeV. 

\begin{figure}[t!]
    \centering
     \includegraphics[width=0.65\textwidth]{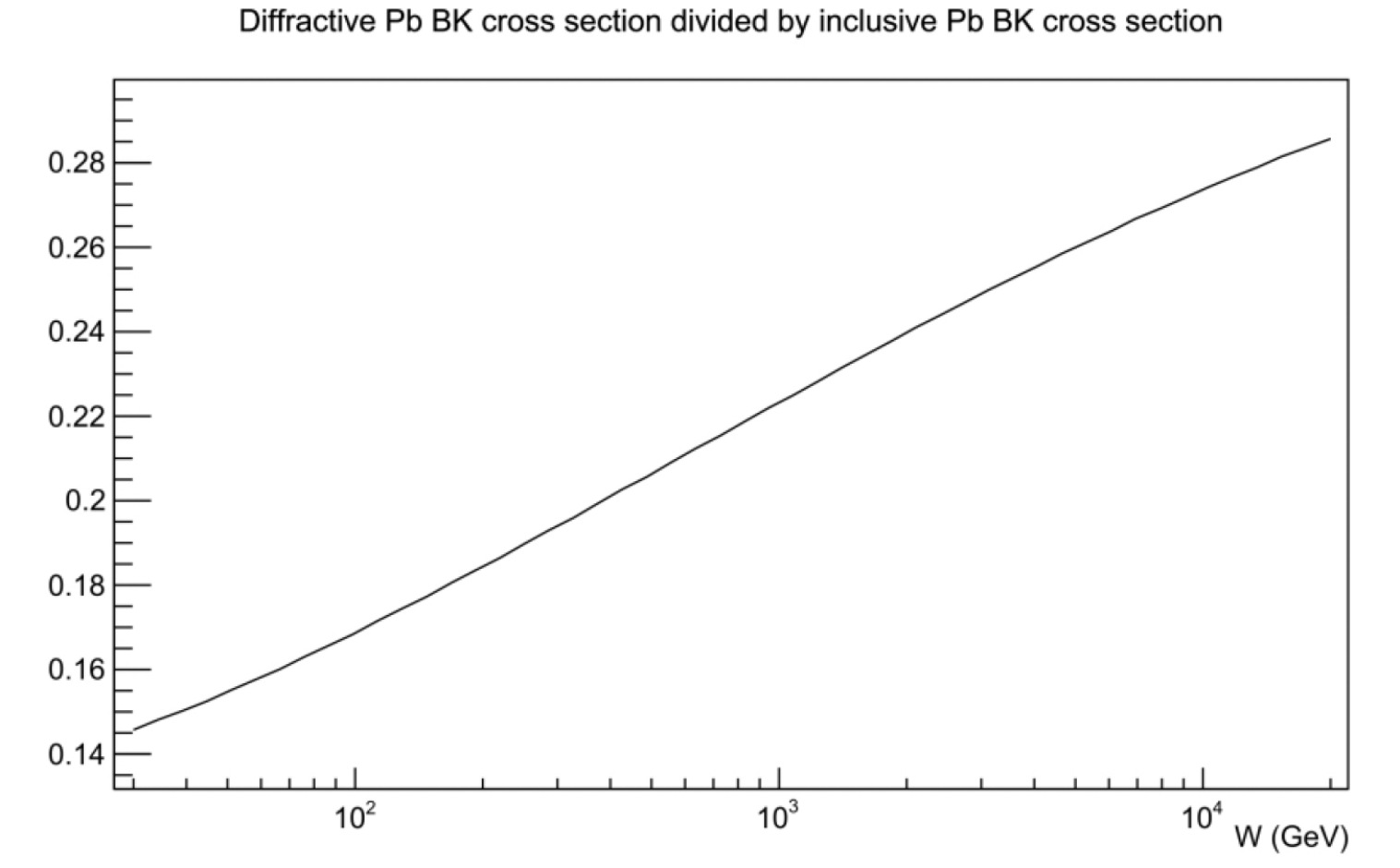}
     \caption{Ratio of the diffractive to the inclusive $c \bar{c}$ production cross sections as predicted by the BK evolution for $\gamma$p (left) and $\gamma$Pb (right) interactions.
        }
    \label{ratio}
\end{figure}

The ratio of the diffractive to the inclusive $c \bar{c}$ production cross sections as predicted by the BK evolution for $\gamma$p (left) and $\gamma$Pb (right) interactions is shown in Fig.~\ref{ratio}. In the case of protons, we predict a ratio of about 12\% which is compatible to what was observed at HERA, and in the case of Pb, we predict a ratio of about 21\%. CMS accumulated Pb Pb interaction data in 2025 that will allow this measurement to be performed (for previous data taking, a positive signal in the zero degree calorimeter was requested at trigger level, removing the diffractive events when Pb is intact after interaction). Such a high percentage of diffractive events will be a clean test of CGC like models and will probably allow to distinguish it from the nuclear PDFs and shadowing approaaches which predict a lower fraction of diffractive events.

Measuring UPC vector meson and charm/beauty production at the LHC and the EIC for different ion interacttions will thus be fundamental to probe saturation models. The reached kinematical domain is complementary between the LHC and the EIC and will allow to obtain measurements with unprecedented precision over a wide domain in energy. The feasibility of running both machines with different ions (He, O, Ar, Ne, Pb, etc...) will allow to study the path towards saturation regime which will appear for heavier ions but not for the light ones, while measuring the nuclear PDFs with high precision.

\section{$\gamma \gamma$ physics at the LHC}

\begin{figure}
\centering
\includegraphics[width=0.4\textwidth]{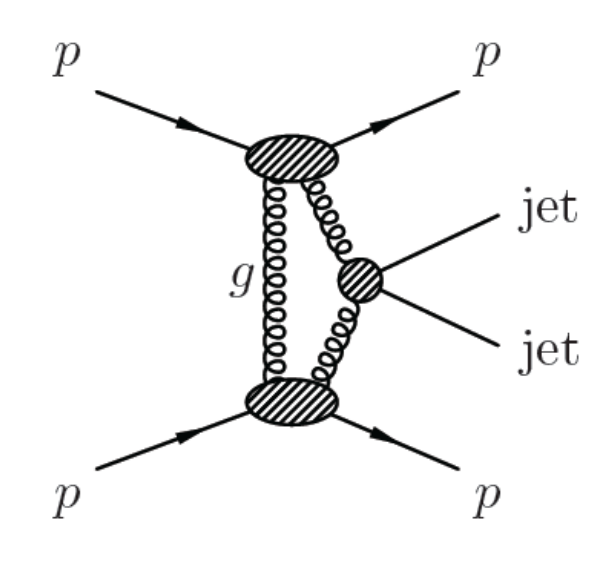}
\includegraphics[width=0.4\textwidth]{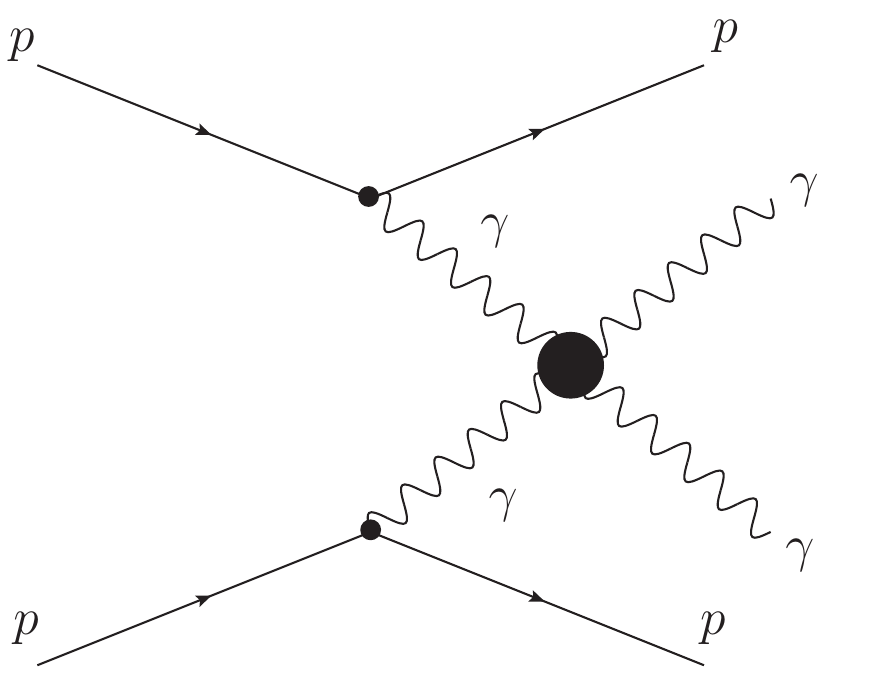}
\caption{Exclusive dijet production via gluon exchanges (QCD process) processes (left) and of diphotons via photon exchanges (QED process,right).}
\label{fig3b}
\end{figure}

In this section, we will describe the reach in beyond standard model physics (especially the production of axion-like particles) when one considers the LHC as a $\gamma \gamma$ collider. The diagrams for exclusive productions via gluon or photon exchanges are shown in Fig.~\ref{fig3b}. The QCD diagram dominates at low diffractive masses (dijet masses for instance) up to masses of about 120 GeV. This allows to study the exclusive production of low mass particles such as pion pair or glueballs in a very clean environment (one can produce glueballs that decay into $\rho \rho$, $\Phi \Phi$, $K^*K^*$, $\omega \omega$ for instance), in addition to more standard studies of pomeron structure in terms of quarks and gluons using diffractive production of jets, photon and jet, $W$ asymmetries, etc~\cite{usdiff}. In this section, we will focus on the exclusive production of $\gamma$, $W$, $Z$ pairs at high masses which is dominated by photon exchanges. We thus consider the LHC as a $\gamma \gamma$ collider. Most of the results in this section assume that we measure the intact protons after collision in the Precision Proton Spectrometer (PPS)~\cite{pps} by the CMS and TOTEM collaborations or the ATLAS Forward Proton (AFP) by the ATLAS collaboration~\cite{afp}, that leads to a good acceptance in diffractive mass above about 450 GeV.

\begin{figure}
\centering
\includegraphics[width=0.95\textwidth]{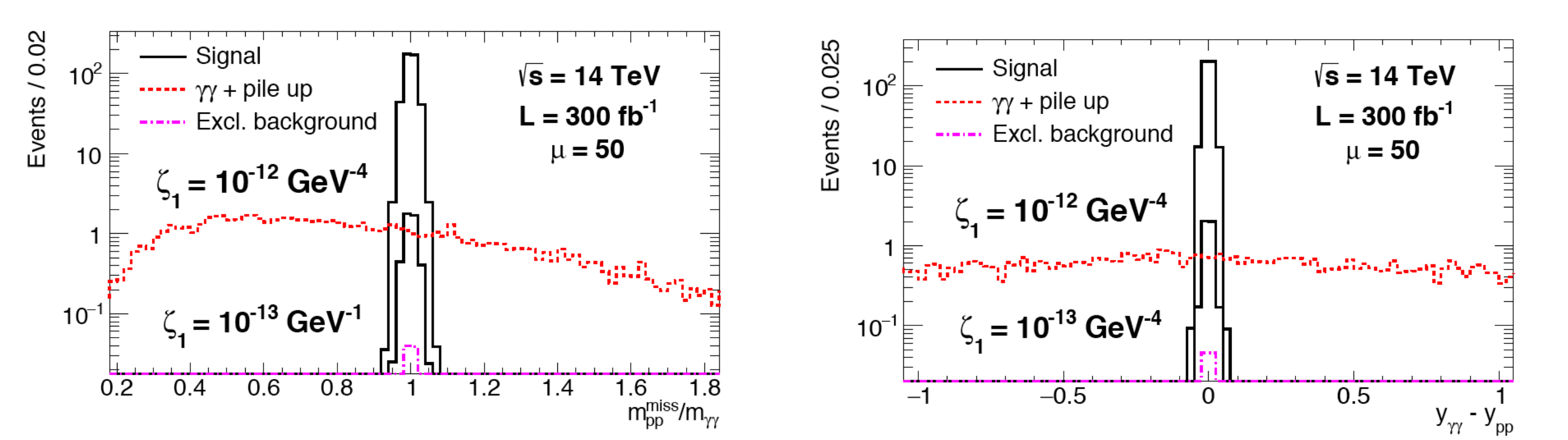}
\caption{Diffractive mass ratio and difference in rapidity using either the diphoton information from CMS or the proton taggings for signal in black and pile up background in red.}
\label{fig4}
\end{figure}

\begin{figure}
\centering
\includegraphics[width=0.45\textwidth]{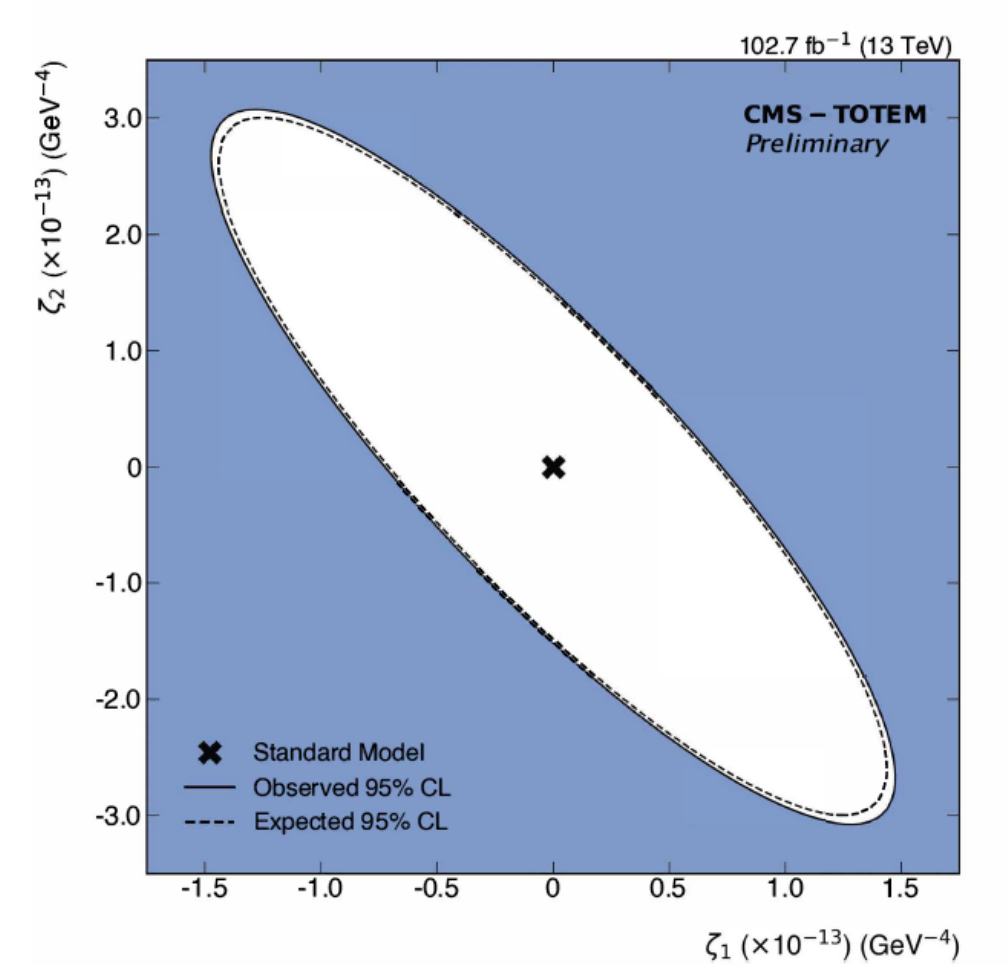}
\includegraphics[width=0.45\textwidth]{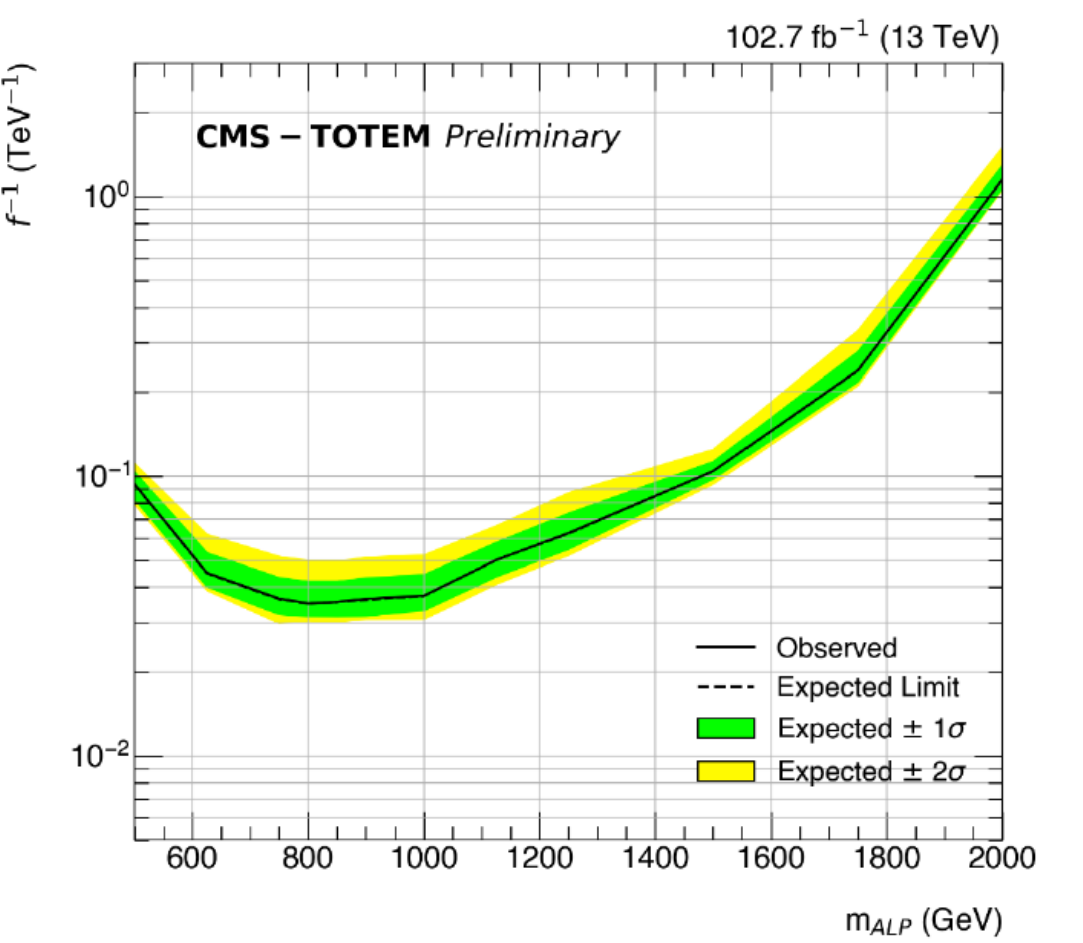}
\caption{Left: limits in the ($\zeta_1$, $\zeta_2$) plane for quartic anomalous $\gamma \gamma \gamma \gamma$ coupling. Right: Limits in the coupling versus mass plane for ALP production.}
\label{fig4b}
\end{figure}

In Fig.~\ref{fig3b}, right, we display the diagram of exclusive $\gamma \gamma$ via photon exchanges. Similar diagrams also include the production of $WW$, $ZZ$, $t \bar{t}$ and $\gamma Z$. As an example, the SM exclusive production of $WW$ is $\sigma_{WW}=95.6$ fb, and $\sigma_{WW}(W=M_X>1 TeV)=5.9$ fb, which shows that many events will be produced at high mass at the LHC (the typical luminosity for Run 3 is 300 fb$^{-1}$ and will increase to 3000 fb$^{-1}$ at the high luminosity LHC).  The exclusive production of $WW$, $ZZ$, $\gamma \gamma$, $t \bar{t}$, $\gamma Z$ will thus be sensitive to different quartic anomalous couplings involving photons, and represents a stringent test. of QED~\cite{diphoton,diphotonb}.

We will first consider the exclusive $\gamma \gamma$ production via photon exchanges as an example. The SM cross section for such a process for diphoton masses above 450 GeV (in the PPS acceptance) is very small and leads to a  negligible number of events. Anomalous quartic $\gamma \gamma \gamma \gamma$ can appear via loops of new particles or resonances decaying into two photons such as axion-like particles (ALPs). 
We introduce two effective operators at low energies in the Lagrangian due to quartic anomalous couplings \begin{eqnarray}
L_{4 \gamma} = \zeta_1^{\gamma} F_{\mu \nu}F^{\mu \nu} F_{\rho \sigma} F^{\rho \sigma} + \zeta_2^{\gamma} F_{\mu \nu}F^{\nu \rho} F_{\rho \lambda} F^{\lambda \mu} 
\end{eqnarray}
and $\gamma \gamma \gamma \gamma$ couplings can be modified in a model
independent way by loops of heavy charged particles
\begin{eqnarray}
\zeta_1 = \alpha_{em}^2 Q^4 m^{-4} N c_{1,s}  
\end{eqnarray}
where the coupling depends only on $Q^4 m^{-4}$ (charge and mass of the
charged particle) and on spin, $c_{1,s}$ which 
leads to $\zeta_1$ of the order of 10$^{-14}$-10$^{-13}$ GeV$^{-4}$. 
$\zeta_1$ can also be modified by neutral particles
at tree level (extensions of the SM including scalar,
pseudo-scalar, and spin-2 resonances that couple to the photon),
\begin{eqnarray}
 \zeta_1 =
(f_s m)^{-2} d_{1,s}
\end{eqnarray}
where $f_s$ is the $\gamma \gamma X$ 
coupling of the new particle to the
photon, and $d_{1,s}$ depends on the spin of the particle. For instance, 2 TeV
dilatons lead to $\zeta_1 \sim$ 10$^{-13}$ GeV$^{-4}$. 
In Ref.~\cite{diphotonb}, we performed an exact calculation and compared it with the usual effective Lagrangian approach and results were found to be similar.

The largest background to be considered to anomalous $\gamma \gamma \gamma \gamma$ events is the diphoton inclusive production (where the protons are destroyed in the final state) and intact protons detected in PPS originate from pile up events. The number of pile up events, or additional proton interactions within one bunch crossing, can reach 50 and will increase to 200 at the high luminosity LHC. This means that we detect intact protons originating from pile up every few bunch crossings. The fact that we measure all particles in the final state (the two $\gamma$s and the two intact protons) allows rejecting pile up background in a very efficient way as shown in Fig.~\ref{fig4}~\cite{diphoton}. For signal, the ratio of the diproton missing mass and the diphoton mass and the difference between the diphoton and diproton rapidity should peak respectively around 1. and 0. within detector resolution because of kinematics conservation. For pile up, we expect a flat distribution since the diphotons and the diprotons do not originate from the same interaction. Requesting the matching between diphoton and diproton information allows obtaining a negligible background for 300 fb$^{-1}$. 

\begin{figure}
\centering
\includegraphics[width=0.8\textwidth]{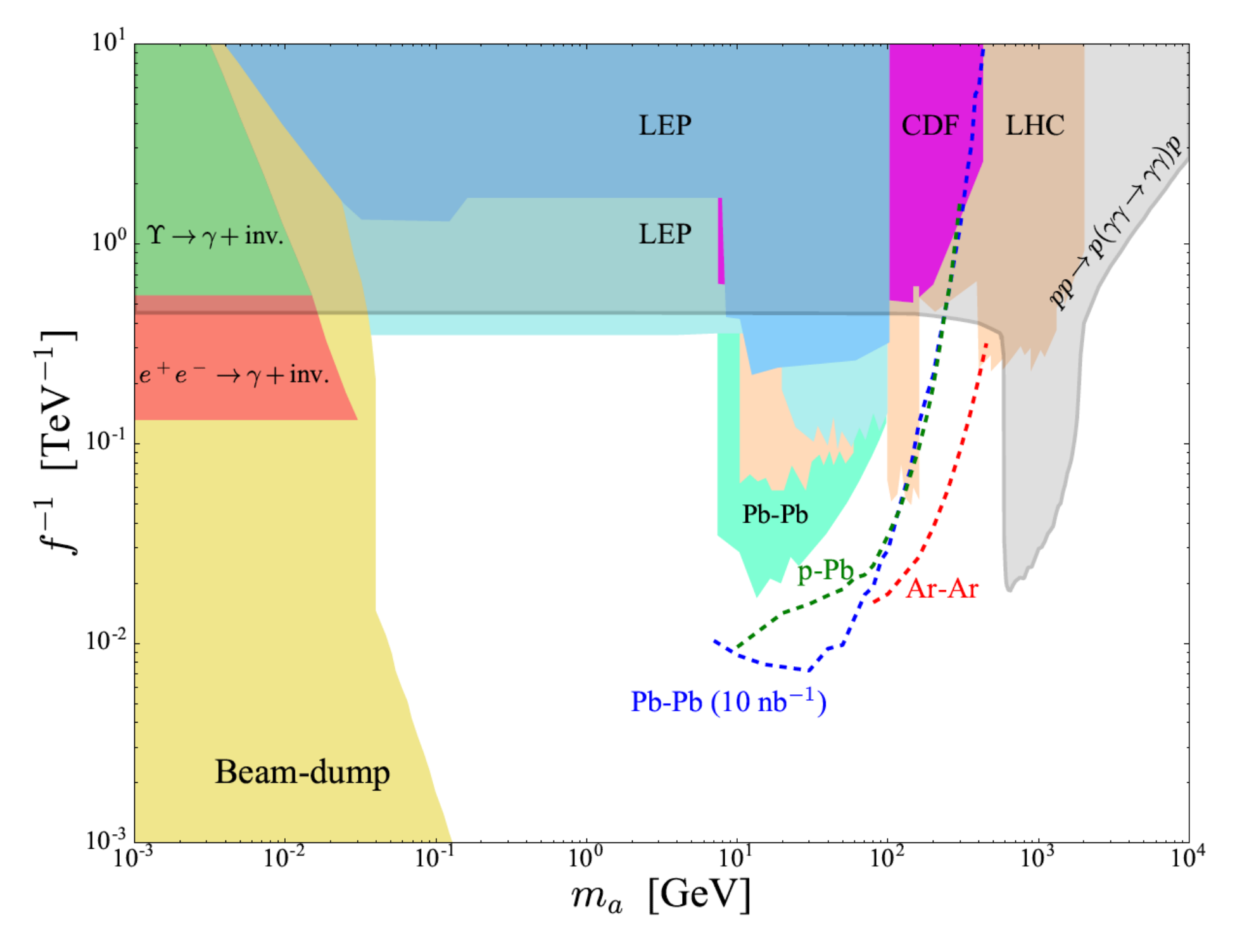}
\caption{Sensitivity plot in the coupling versus mass plane for ALP production for $pp$ and heavy ion interactions.}
\label{fig5}
\end{figure}

 Recently, the CMS collaboration performed the search for exclusive diphoton production requesting two back-to-back photons at high diphoton mass ($m_{\gamma \gamma}>350$ GeV), matching in rapidity and mass between the diphoton and diproton informations. This led to the first limits on quartic photon anomalous couplings, $|\zeta_1|<2.9~10^{-13}$ GeV$^{-4}$, $|\zeta_2|<6.~10^{-13}$ GeV$^{-4}$ with about 10 fb$^{-1}$~\cite{photoncms}.
 The limit was then updated with 102.7 fb$^{-1}$ as $|\zeta_1|<7.3~10^{-14}$ GeV$^{-4}$, $|\zeta_2|<1.5~10^{-13}$ GeV$^{-4}$~\cite{photoncms}, as shown in Fig.~\ref{fig4b}, left. These results can be reinterpreted as a limit on ALPs at high masses and the results are shown in Fig.~\ref{fig4b}, right, as limits in the coupling versus mass plane. 
 In Fig.~\ref{fig5}, we also display the sensitivities projected for 300 fb$^{-1}$ of luminosity~\cite{alp}.  The sensitivity for $pp$ collisions is shown in a greyish area at high ALP mass and we see that we gain about two orders of magnitude for ALP masses of about 1 TeV, and we reach an uncovered region at higher massses. It is also worth noting that the
 production of ALPs via photon exchanges in heavy ion runs is complementary to the $pp$ interactions and allows to cover the intermediary region in ALP masses between a few GeV and 1 TeV. 
Heavy ion runs show much lower luminosity than $pp$ runs but the cross section is increased by $Z^4$~\cite{alp}.

In Ref.~\cite{exclgammaz},  we also studied the sensitivity to quartic anomalous $\gamma \gamma \gamma Z$ coupling, considering the leptonic and hadronic $Z$ boson decays.  The fact that we can
control the background using the mass and rapidiy matching technique allows us to
look in both channels with negligible background. This
leads to a very good sensitivity to $\gamma \gamma \gamma Z$ couplings, about three orders of magnitude better than the standard search at the LHC (looking for the $Z$ boson decay into three photons is very challenging in a high pile up environment). The advantage of our method is also that the 
sensitivity to anomalous couplings is performed in a
model independent way and the anomalous coupling  can be due to wide or narrow resonances or loops of new
particles as a threshold effect.

\begin{figure}
\centering
\includegraphics[width=0.45\textwidth]{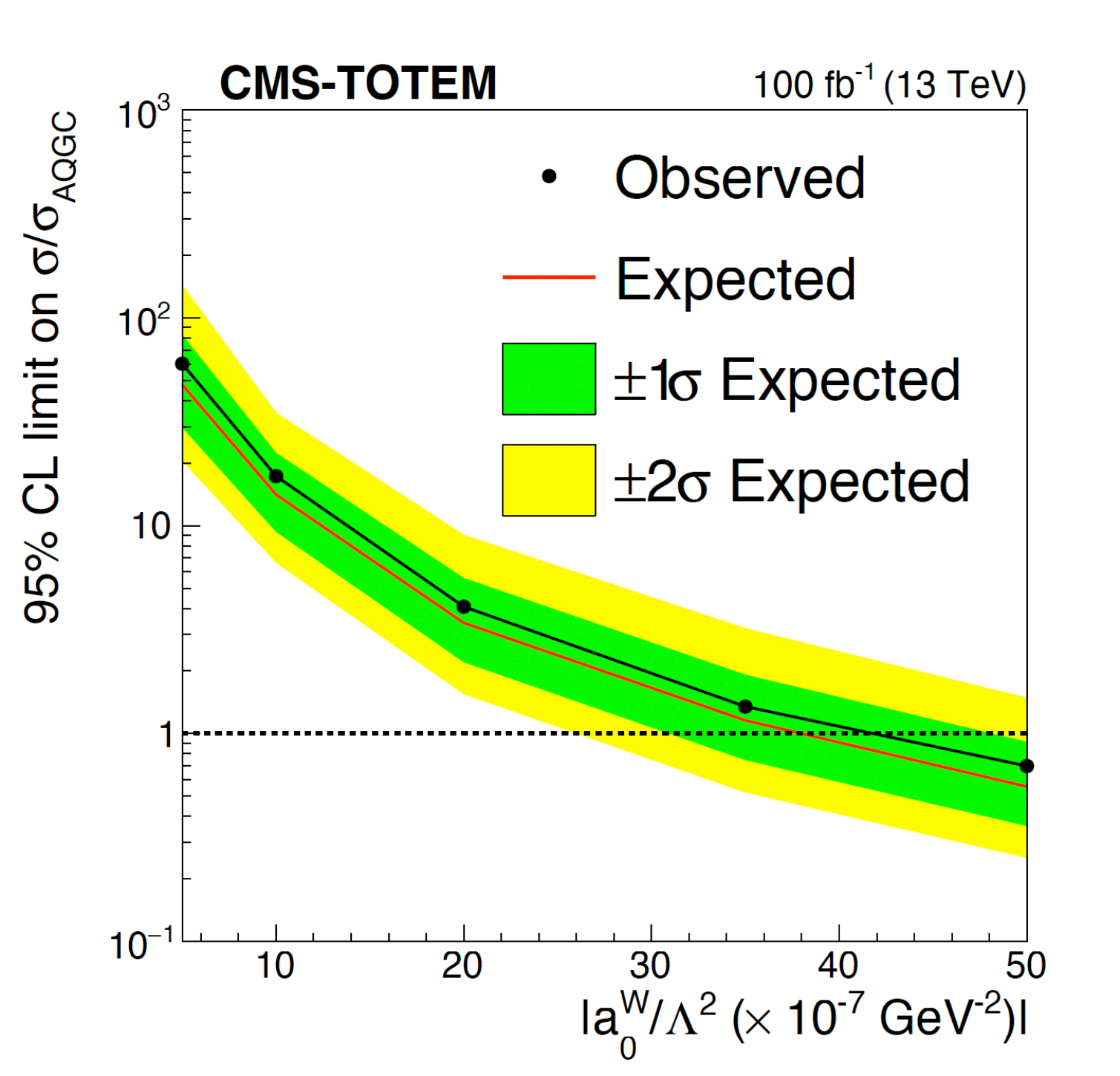}
\includegraphics[width=0.49\textwidth]{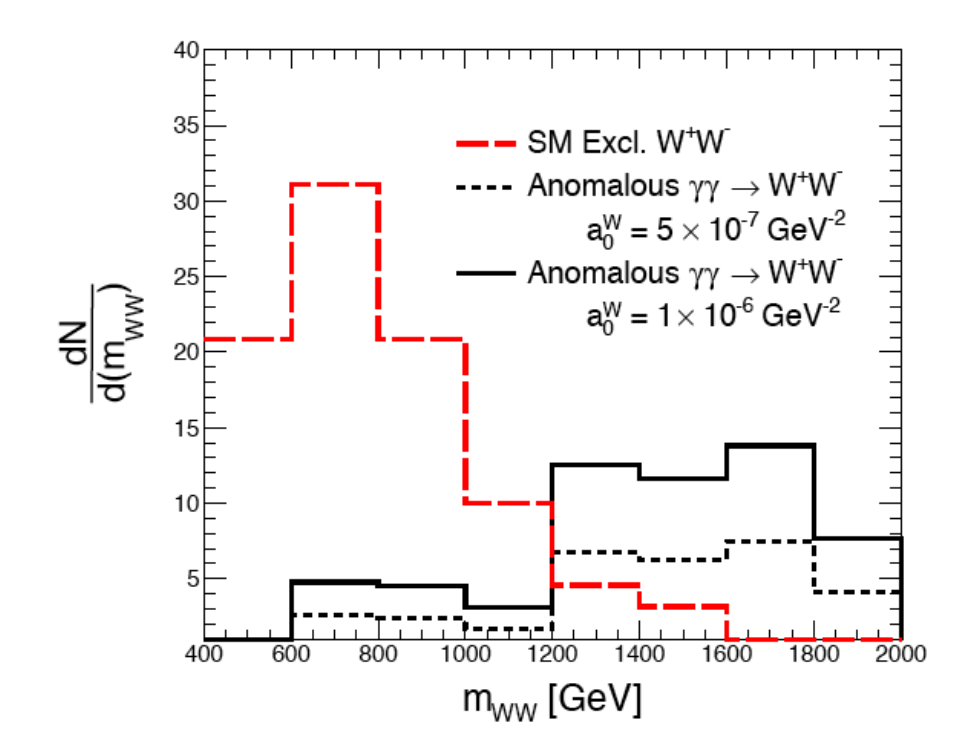}
\caption{Left: limits on $a_0^W$ for quartic anomalous $\gamma \gamma WW$ coupling. Right: Mass distribution for exclusive $WW$ production for SM (in red dashed line) and for two values of quartic anomalous $\gamma \gamma WW$ coupling (in black full and dotted lines).}
\label{fig5b}
\end{figure}

The CMS collaboration also looked for quartic  $\gamma \gamma WW$ and $\gamma \gamma ZZ$ anomalous couplings studying the case where the $W$ or $Z$ bosons decay fully hadronically. This is because the anomalous production of $WW/ZZ$ events dominates at high mass with a rather low cross section and the branching ratio into hadrons is higher. Two ``fat" jets of radius 0.8 were selected with a  transverse momentum $p_T>200$ GeV,  a dijet mass 1126$<m_{jj}<$2500 GeV, and the jets are requested to be back-to-back ($|1-\phi_{jj}/\pi|<0.01$). As usual, the
signal region is defined by the matching between the central $WW$ system and the proton information. The results are shown in Fig.~\ref{fig5b}, left, and the 
limits on SM cross section are $\sigma_{WW}<67$ fb, $\sigma_{ZZ}<43$ fb for $0.04<\xi<0.2$~\cite{cmsww}. The  
new limits on quartic anomalous couplings with events violating unitarity removed are $a_0^W/\Lambda^2 < 4.3~10^{-6}$ GeV$^{-2}$, $a_C^W/\Lambda^2 < 1.6~10^{-5}$ GeV$^{-2}$, $a_0^Z/\Lambda^2 < 0.9~10^{-5}$ GeV$^{-2}$, $a_C^Z/\Lambda^2 < 4.~10^{-5}$ GeV$^{-2}$ with 52.9 fb$^{-1}$. In Ref.~\cite{exclww}, we also studied the possible observation of exclusive $WW$ production with 300 fb$^{-1}$ of data. The
SM contribution appears at lower $WW$ masses, as shown in Fig.~\ref{fig5b}, right, compared to anomalous couplings. This is why one needs to consider
purely leptonic  channels for $W$ decays (the dijet background being too high at low masses for hadronic channels).
The SM prediction on exclusive WW (leptonic decays) after selection is about 50 events for 300 fb$^{-1}$  for a background estimate of 2 events.

The CMS collaboration also performed the search for exclusive $\gamma \gamma t \bar{t}$ processes in leptonic and  semi-leptonic decays of top quarks with 29.4 fb$^{-1}$ that led to a limit of $\sigma_{excl}. < 0.59$ pb~\cite{cmsexclttbar}.  A study concerning the sensitivity to $\gamma \gamma t \bar{t}$  quartic anomalous couplings was performed in Ref.~\cite{exclttbar} using a selection based on matching between the $pp$ and $t \bar{t}$ information and on
fast timing detectors to further suppress the pile up background, and the sensitivity can reach 7 10$^{-12}$ GeV$^{-4}$ for 300 fb$^{-1}$ of $pp$ collisions using timing detectors with a resolution of about 20 ps.

In general, tagging protons at the LHC and matching the information of the tagged proton and the objects measured in the main CMS detector allow to increase the sensitivity to $\gamma \gamma \gamma \gamma$, $\gamma \gamma  WW$, $\gamma \gamma ZZ$, $\gamma \gamma \gamma Z$, $\gamma \gamma t \bar{t}$ quartic anomalous couplings by two or three orders of magnitude compared to more standard measurements at the LHC.

\section{Conclusion}

In this review, we presented first the discovery of the odderon by comparing D0 $p \bar{p}$ and TOTEM $pp$ elastic interactions. We then discussed the observation of gluon resummation effects at low $x$ especially in the measurement of gap between jets and the possible observation of saturation in PbPb collision via the measurement of $J/\Psi$ mesons, $c \bar{c}$ and $b \bar{b}$ production cross sections. We finished the report by discussing the sensitivity to quartic anomalous couplings and to the existence of ALPs by measuring intact protons at the LHC at high luminosity.

\section*{Acknowledgments}
Many of these results originate from common work with long term collaborators and I would like to thank all of them.
I also thank Robi Peschanski for a careful reading of the manuscript.


\end{document}